\documentclass[sigconf]{acmart}
\usepackage{pgfplots}
\usepackage{booktabs}
\usepackage{amsmath}
\usepackage{tikz}
\usepackage{multirow}
\usepackage{subfigure}

\setcopyright{none}

\acmDOI{}

\acmISBN{}

\acmConference[Working paper]{}{Wisconsin}{USA}

\acmBooktitle{}
\copyrightyear{September, 2023}

\begin{document}
\title{American Option Pricing using Self-Attention GRU and Shapley Value Interpretation}

\author{Yanhui Shen}
\affiliation{University of Wisconsin-Madison}
\email{yshen272@wisc.edu}


\begin{abstract}
Options, serving as a crucial financial instrument, are used by investors to manage and mitigate their investment risks within the securities market. 
Precisely predicting the present price of an option enables investors to make informed and efficient decisions.
In this paper, we propose a machine learning method for forecasting the prices of SPY (ETF) option based on gated recurrent unit (GRU) and self-attention mechanism. 
We first partitioned the raw dataset into 15 subsets according to moneyness and days to maturity criteria. For each subset, we matched the corresponding U.S. government bond rates and Implied Volatility Indices. This segmentation allows for a more insightful exploration of the impacts of risk-free rates and underlying volatility on option pricing. 
Next, we built four different machine learning models, including multilayer perceptron (MLP), long short-term memory (LSTM), self-attention LSTM, and self-attention GRU in comparison to the traditional binomial model.
The empirical result shows that self-attention GRU with historical data outperforms other models due to its ability to capture complex temporal dependencies and leverage the contextual information embedded in the historical data.
Finally, in order to unveil the "black box" of artificial intelligence, we employed the SHapley Additive exPlanations (SHAP) method to interpret and analyze the prediction results of the self-attention GRU model with historical data. This provides insights into the significance and contributions of different input features on the pricing of American-style options.
\end{abstract}

\keywords{American option, Option pricing, Binomial tree, Deep learning, Multilayer perception, LSTM neural network, Gated recurrent unit, Self-attention, Interpretable machine learning, Shapley values.}

\maketitle

\section{Introduction}
Option pricing is a crucial topic within the branch of financial research known as asset pricing, which involves the calculation of fair prices for option contracts. Options can be categorized into two main types based on their exercise time: European options and American options. In 1973, Black and Scholes \cite{black1973pricing}, pioneers in the field of option pricing, introduced the renowned Black-Scholes model that provides closed-form solutions to European options. In comparison, the valuation of American options is more challenging due to the possibility of early exercise and flexible exercise schedules. To address these issues, numerical methods are often required to approximate American option prices while considering optimal stopping strategies and market dynamics. The prevalent conventional approaches in the literature for valuing American options include binomial tree models \cite{cox1979option,boyle1986option,kim2016multi}, numerical partial differential equation (PDE) techniques \cite{brennan1977valuation, ikonen2004operator, nielsen2002penalty, zhao2007compact}, and Monte Carlo (MC) simulations \cite{bossaerts1989simulation, tilley1993valuing, broadie1997enhanced, broadie1997pricing, longstaff2001valuing}. However, most traditional methods for pricing American options are constrained by rigid assumptions and can suffer from the curse of dimensionality. This can easily lead to inaccurate predicted results and an exponential increase in computational time under high-dimensional data scenarios. As computer technology continues to evolve, the application of data science and machine learning in the financial sector is growing. The resulting innovations in option pricing methods are profound and cannot be overlooked.

The concept of neural networks was first proposed by McCulloch and Pitts \cite{mcculloch1943logical} in 1943, but at that time, due to limitations in computer technology, it did not receive widespread attention and applications. In recent years, propelled by advancements in computer hardware and software, neural networks have found extensive use in fields such as computer vision (CV), natural language processing (NLP), automatic speech recognition (ASR), finance, healthcare, autonomous driving, and more. Malliaris and Salchenberger \cite{malliaris1993neural} are the forerunners to employ neural networks in the domain of option pricing. 
Subsequently, an expanding group of researchers started to focus on the advantages of neural network models over conventional ones in European option pricing and have made contributions \cite{hutchinson1994nonparametric,lajbcygier1997improved,anders1998improving,bennell2004black,liang2009improving,wang2009nonlinear,garcia2000pricing,park2014parametric}. 
The neural network has the capability to approximate nonlinear models and handle high-dimensional data, which makes it equally suitable for pricing American options \cite{kelly1994valuing,keber1999option,morelli2004pricing,pires2004american,jang2019generative,kohler2010pricing}. 
Initially, the application of neural networks in financial research was confined to basic single-layer structures. With the development of deep learning algorithms, scholars begin to explore the utilization of complex neural network architectures, such as convolutional neural network (CNN) and long short-term memory (LSTM) \cite{siami2018comparison,hiransha2018nse,hirsa2019supervised,livieris2020cnn,becker2019deep}. However, articles that implement deep learning for American option pricing are still relatively scarce, indicating that there is substantial untapped potential for growth in this interdisciplinary field. 

Artificial intelligence algorithms have progressed from basic decision rules to advanced deep learning techniques, continuously improving accuracy and computational efficiency. On the other hand, the interpretability of the models has progressively diminished. This dilemma of the trade-off between accuracy and interpretability has given rise to the field of explainable artificial intelligence (XAI). 
In April 2018, the European Commission issued a communication about AI in Europe to key European bodies, such as the European Parliament and the European Council. 
The communication highlights the importance of exploring AI system explainability to bolster public confidence in AI technologies. 
The government requirements and public trust demands have led to the emergence of common interpretability methods, including partial dependence plot (PDP), accumulated local effects (ALE), local interpretable model-agnostic explanations (LIME), and SHapley Additive exPlanations (SHAP). 
Among the limited literature exploring American option pricing within deep learning, few have focused on interpretability.  This paper identifies the gap and attempts to address this deficiency, shedding light on an overlooked area. 

This paper introduces four distinct deep learning models for pricing American options: multilayer perception (MLP), long short-term memory (LSTM), self-attention LSTM and self-attention gated recurrent unit (GRU). We compare these four neural network pricing models with the classical binomial tree model based on transactions of the SPY (ETF) call option from January 2020 to December 2022. The empirical results show the attention-GRU model demonstrates the best performance and remarkably reduces the computational time required in the conventional binomial tree model. Then, in order to make the best model understandable and open the "black box" of neural networks, we adopt the SHapley Additive exPlanations (SHAP), a technique for machine learning interpretability. The concept of Shapley values was originally introduced by Shapley \cite{shapley1953value} in 1953 to resolve the issue of fair allocation in game theory. With the evolution of eXplainable Artificial Intelligence (XAI), Lundberg et al. \cite{lundberg2017unified} offered the SHAP algorithm in a unified framework that can be used to explain all machine learning models. To the best of our knowledge, we find that there is no existing research applying this interpretable method to the pricing of American options.

The contribution of this paper includes four aspects. 
First, we present various deep learning models for valuing American options. There are not many articles in this specific interdisciplinary field, and few studies have incorporated attention mechanisms into LSTM or GRU neural networks. 
Second, we match the categorized subsets of data based on moneyness and days to maturity with corresponding U.S. government bond rates and Implied Volatility Indices. By doing this, we can more effectively analyze the effects of underlying volatility and interest rates on option prices, relaxing the fixed assumptions of the two inputs in traditional models. Additionally, we innovatively select different Implied Volatility Indices as features in our neural network models to match options with varying expiration dates. This enables a better approximation of the underlying return volatility across distinct future time periods. 
Third, in terms of feature engineering, we take into consideration the impact of historical call prices on the option price that needs to be predicted. The experiment result indeed proves that historical call prices have significant impacts across all categories of options, especially for the at-the-money type. We also include spot price, strike, and moneyness as features in our deep learning model training, allowing us to analyze the influence of the correlation between spot price and strike on the predicted option price. The last, to the best of our knowledge, this article is the first to utilize Shapley value as an interpretive tool for pricing American options. The research indicates that SHAP is user-friendly, visually robust, and capable of providing highly intuitive explanations on a global scale. The SHAP tenique bridges the gap created by highly accurate yet less interpretable machine learning models, thereby bolstering the overall comprehensiveness and credibility of our analysis.

The remainder of the paper is organized as follows. Section 2 provides a brief review of the traditional pricing methods, machine learning methods, and interpretable machine learning methods in option pricing fields. Section 3 explains the mathematical and algorithmic theories of all models this paper adopts. Section 4 introduces the data and its pre-processing process. Section 5 displays the whole experiment process and analyzes the detailed results. Section 6 concludes the paper and suggests several directions for future research.

\section{Literature review}
Since the seminal Black–Scholes model (B-S model) proposed by Black and Scholes \cite{black1973pricing} in 1973, option pricing as the canonical topic in both the finance academia and the financial market has been deeply studied for a long time. To delve into the evolution of option pricing methodologies, the existing literature can be generally classified into three groups: traditional pricing methods, machine learning methods, and interpretable machine learning methods.
\subsection{Traditional pricing methods}
American-type options can be exercised at any time before expiration, so this early exercise privilege turns American option pricing into an optimal stopping problem that is closely associated with personal preference and market uncertainty. The B-S model provides the pioneering analytical solution to pricing European options, but Merton \cite{merton1973theory} showed the difficulty in obtaining closed-form formulas of American styles with the B-S model framework.

American options are path-dependent which makes the valuation more complex and conventional methods focus extensively on computationally intensive numerical methods. The predominant pricing approaches encompass binomial tree models, numerical partial differential equation (PDE) techniques, and Monte Carlo (MC) simulations.

Cox et al. \cite{cox1979option} was the first to propose the celebrated binomial tree model (Cox-Ross-Rubinstein model), which replicates the B-S model in discrete time. The binomial tree approach has gained widespread adoption in the field of finance due to its versatility and ease of implementation. However, it proves ineffective when dealing with stochastic volatility. In 1986, Boyle \cite{boyle1986option} incorporated a middle price path into the price tree to improve the binomial tree method. The derived trinomial tree can converge faster to actual option values compared to the binomial tree. In 2016, Kim et al. \cite{kim2016multi} constructed a new binomial tree model fitting all moments to the approximated geometric Brownian motion that generalizes the classical Cox-Ross-Rubinstein model, resolving a discontinuity problem in American option pricing. The binomial model provides transparency into the underlying prices in different paths and early exercise moments, which makes the pricing process interpretable and intuitive. Yet, the convergence of interval oscillations in the binomial tree method makes it challenging to apply in high-dimensional data. As the number of time steps increases, the number of tree branches will exponentially explode,  leading to reduced computational efficiency. Besides, the binomial tree method relies on a set of predefined assumptions and parameters, which may not accurately capture the complexity of real-world financial data. 

Another prevailing methodology for pricing American options is the partial differential equation (PDE) method which regards the core optimal stopping problem of American options as a PDE boundary condition. In 1977, Brennan et al. \cite{brennan1977valuation} employed a boundary projection method to price American put options and applied it to the put contracts traded in the New York dealer market. Later, Ikonen et al. \cite{ikonen2004operator} proposed operator splitting methods based on the Crank-Nicolson method \cite{crank1947practical} and the two-step backward differentiation formula \cite{wanner1996solving}, and the numerical experiments show that their methods are much more efficient than the boundary projection method. Recently, there are other derived techniques for pricing American options such as penalty and front-fixing method \cite{nielsen2002penalty}, compact finite difference method \cite{zhao2007compact}, and two dimensional partial differential integral equation (PDIE) model \cite{yousuf2023partial}. However, solving PDE problems numerically for a wide range of option contracts can require substantial computational resources and time, particularly for high-dimensional problems or intricate option structures.

In addition to the aforementioned traditional approaches, the statistical simulation methods known as Monte Carlo (MC) methods are widely applied in American option pricing. MC methods have demonstrated high flexibility in pricing derivatives under different volatility dynamics, especially in the case of intricate path-dependent payoff structures. Bossaerts \cite{bossaerts1989simulation} and Tilley \cite{tilley1993valuing} are the trailblazers to exploit simulation in American option pricing. However, initial MC methods failed to effectively deal with high-dimensional problems. In 1997, Broadie et al. \cite{broadie1997enhanced} suggested "pruning" the trees by reducing branches whenever possible, thus lessening the simulated time and allowing for faster convergence of the estimates than that in the Tilley method. In the same year, Broadie et al. \cite{broadie1997pricing,broadie1997sotchastic} introduced two stochastic mesh approaches for pricing high-dimensional American options. Subsequently, Longstaff and Schwartz \cite{longstaff2001valuing} put forward an innovative technique for pricing American options using simulation, which they referred to as the least-squares Monte Carlo (LSM) approach. The LSM method was proven to obtain accurate and computationally efficient predictions for various path-dependent and exotic options. Despite the high flexibility shown in MC methods, the curse of dimensionality becomes prominent as the number of underlying assets and simulated paths increases. This can result in a decline in the precision of predictive outcomes and a sharp rise in computational time.

\subsection{Machine learning methods}
Given the ever-changing nature of financial markets, there is always a pressing need for methods that provide more accurate and efficient pricing than traditional models, in order to effectively avoid arbitrage opportunities.

Ever since the initial contribution of Malliaris and Salchenberger \cite{malliaris1993neural} in 1993, a growing number of scholars have been focusing on the utilization of artificial neural networks (ANNs) in option pricing. 
Hutchinson et al. \cite{hutchinson1994nonparametric} proposed a nonparametric method for estimating the prices of S$\&$P 500 futures call options using learning networks. The authors presented that nonparametric learning network methods can be useful substitutes when parametric methods fail. Lajbcygier et al. \cite{lajbcygier1997improved} employed a hybrid neural network to predict the discrepancy between the conventional option pricing model and real-world observed option prices. They also found greatly improved predictions through the application of bootstrap methods. Anders et al. \cite{anders1998improving} proved that the combination of statistical inference techniques and neural network methods results in superior out-of-sample pricing performance compared to the Black-Scholes model. Bennell et al. \cite{bennell2004black} was the first to price UK options in the framework of neural networks and allow for dividends in the closed-form model. Liang et al. \cite{liang2009improving} improved the option price forecasts with neural networks and support vector regressions. Wang \cite{wang2009nonlinear} attempted to integrate GJR–GARCH volatility into ANN option-pricing model, which shows higher predictability than other volatility approaches. Garcia et al. \cite{garcia2000pricing} provided that a feedforward neural network with the homogeneous option pricing function reduces the out-of-sample mean squared prediction error. Park et al. \cite{park2014parametric} revealed that non-parametric machine learning methods significantly outperform parametric methods on both in-sample pricing and out-of-sample pricing based on the KOSPI 200 index options from January 2001 to December 2010. Ivașcu \cite{ivașcu2021option} examined the effectiveness of machine learning techniques (i.e., Support Vector Regressions, Genetic Algorithms, and Decision Tree methods) versus classical approaches like Black-Scholes and Corrado-Su \cite{corrado1996skewness} models on European call option pricing. The experiments indicate machine learning algorithms decisively beat traditional models.

The literature discussed earlier primarily focuses on European options. However, American options, due to their inherent ability for early exercise, introduce greater uncertainty and complexity  when estimating prices. This scenario makes machine learning pricing a more suitable alternative than traditional pricing methods since machine learning methods exhibit a natural advantage in handling complex nonlinear relationships and high-dimensional data. In 1994, Kelly et al. \cite{kelly1994valuing} took the initiative to price American put options using a neural network approach and construct hedged portfolios. Keber \cite{keber1999option} employed the genetic programming methodology to value American put option prices on non-dividend yielding stocks and proved the genetic method surpasses conventional approximate analytical formulas. Morelli et al. \cite{morelli2004pricing} compared two kinds of neural networks (i.e., multi-layer perceptrons and radial basis functions) in American option pricing. The results show that radial basis functions can be trained quickly but have limited performance, while the performance of multi-layer perceptrons is superior and robust although the training time is lengthy. Pires et al. \cite{pires2004american} indicated that the support vector machine approach demonstrates better prediction performance than the multi-layer perceptron neural network in the context of American option pricing. Li et al. \cite{li2009learning} studied reinforcement learning (RL) methods to resolve the challenge of early exercise for American options. Their empirical research presents that the RL methods excel over the classical  Longstaff-Schwartz \cite{longstaff2001valuing} method on both real and simulated datasets. On a similar note, Lin et al. \cite{lin2021american} extended the Longstaff-Schwartz \cite{longstaff2001valuing} method using machine learning methods such as support vector regression and regression trees. Jang et al. \cite{jang2019generative} introduced a generative Bayesian learning model that integrates a risk-neutral pricing framework. Their experiment on predicting S$\&$P 100 American put option prices indicates the neural network model achieves the most favorable outcomes. Gaspar et al. \cite{gaspar2020neural} designed two neural network models to price the American options of four large U.S. firms and utilized the Least-Squares Monte Carlo (LSM) method for comparison. It turns out that all neural network methods present lower errors than the LSM method. Kohler et al. \cite{kohler2010pricing} investigated the least squares neural networks regression method to estimate the American option prices whose underlying is simulated by Monte Carlo methods. Li \cite{li2022iteration} provided an iteration algorithm to price American options based on reinforcement learning theories, bringing about good performances in the convergence analysis and the accuracy test. Anderson et al. \cite{anderson2022accelerated} proposed an American option pricing method that combines the classical Heston stochastic volatility model \cite{heston1993closed} and neural network methods. 

In recent years, with the significant enhancement of foundational computational power and the realization of large-scale datasets, the concept of deep learning has garnered increased attention from researchers, and the realm of finance is no exception. The long short-term memory (LSTM) model proposed by Hochreiter and Schmidhuber \cite{hochreiter1997long}, being among the foremost deep learning architectures, possesses a robust ability to analyze time-series data. Siami-Namini et al. \cite{siami2018comparison} applied LSTM to predict financial data series such as NASDAQ composite index (IXIC) and S$\&$P 500 commodity price index (GSPC). The experiments display that LSTM algorithms significantly perform better than traditional methods like ARIMA models. Liang et al. \cite{liang2020forecasting} adopted LSTM neural network to forecast the peer-to-peer (P2P) platform default rate, resulting in higher accurate prediction than other traditional models. Hiransha et al. \cite{hiransha2018nse} employed four kinds of deep learning techniques (i.e., multilayer perceptron, Recurrent Neural Networks, long short-term memory, and convolutional neural networks\footnote{This paper also employs a one-dimensional convolutional neural network model for American option pricing, but finds it less accurate than other deep learning models. Therefore, related results are omitted from the main body.}) to forecast the stock prices from two different stock markets. Among all these methods, the convolutional neural network (CNN) exhibits the highest level of performance. Hirsa et al. \cite{hirsa2019supervised} experimented with different configurations involving varying layer numbers, neurons per layer, and activation functions while employing supervised deep neural networks (DNNs) for option pricing. Their findings suggest that DNNs can notably improve both the efficiency and accuracy of the pricing process. Livieris et al. \cite{livieris2020cnn} conducted predictions on gold price movements by combining LSTM and convolutional neural networks and present that the forecasting results can be boosted substantially in doing so. Chen et al. \cite{chen2021numerical} attempted to solve the Black-Scholes equations by using Laguerre functions as the activation function of neural network models and applied this novel approach to price European options successfully. Becker et al. \cite{becker2019deep} proposed a deep learning method to analyze optimal stopping problems and then tested this technique in the case of Bermudan option pricing. The case study shows the novel deep learning technique produces highly accurate predictions with a very brief computation time.

Apart from the previously mentioned articles that delve into specific methodologies and experiments, there are also valuable literature review papers that can assist us in comprehending the complete evolution of financial research as well as the advancements and innovations in technologies. Atsalakis et al. \cite{atsalakis2009surveying} surveys more than 100 published articles that apply neural network techniques to predict stock markets. Karagozoglu \cite{karagozoglu2022option} presents a thorough review, mapping the progression from the renowned Black-Scholes-Merton model to cutting-edge technologies in the field of option pricing. Ruf et al. \cite{ruf2019neural} focus on the neural networks for option pricing and hedging since the early 1990s. Cavalcante et al. \cite{cavalcante2016computational}, Tkác et al. \cite{tkavc2016artificial} and Riyazahmed \cite{riyazahmed2021neural} provide literature reviews that highlight the substantial assistance machine learning has offered in various financial subdomains. Additionally, there are several books that illustrate diverse practical applications of machine learning approaches in the financial sector, accompanied by easily understandable case studies, serving as valuable resources for learning and reference. Zirilli \cite{zirilli1997financial} primarily discusses the application of neural networks in financial forecasting. Nagel \cite{nagel2021machine} elaborates extensively on the utilization of machine learning in the context of asset pricing. Hull \cite{hull2021machine} presents a clear and succinct overview of the most commonly used machine learning algorithms, particularly underscoring their practical implementations in business contexts.

The extensive literature outlined above underlines the significant contribution of machine learning to the field of finance. As the realm of artificial intelligence continues to evolve, the interdisciplinary fusion of finance and machine learning holds immense untapped potential. As a critical branch of financial research, option pricing has seen numerous scholars begin to apply machine learning methods to European option pricing. In contrast, the utilization of deep learning to price American options remains particularly scarce. Moreover, the inherent "black box" characteristic of artificial intelligence has brought about the problem of interpretability within this limited body of research, thereby eroding the credibility of these deep learning models. Consequently, the investigation of deep learning interpretability for American option pricing becomes increasingly essential and urgent.

\subsection{Interpretable machine learning methods}
As deep learning and other sophisticated black-box models advance, the societal need or regulatory obligations for the interpretability and explainability of machine learning models are growing in prominence \cite{pasquale2015black,doshi2017accountability}. In machine learning, the terminologies of interpretability and explainability have often been employed interchangeably \cite{bibal2016interpretability}. A formal definition of interpretability has yet to be introduced in the literature \cite{bibal2016interpretability,lipton2018mythos}. Intuitively, interpretability can be perceived as the capacity of a model to be understood by its users. In the early stages of machine learning development, straightforward models such as decision trees can be considered interpretable, whereas complex models like support vector machines (SVM) and neural networks (NN) are often seen as lacking interpretability, commonly referred to as "black boxes".

The pressing demand for interpretability in machine learning models has led to the emergence of explainable artificial intelligence (XAI) research. The mainstream methods for achieving interpretability in machine learning primarily include partial dependence plot (PDP), accumulated local effects (ALE), local interpretable model-agnostic explanations (LIME), and SHapley Additive exPlanations (SHAP). In recent years, there has been a growing trend of researchers attempting to apply interpretable machine learning approaches in the field of finance. The PDP method was first proposed by Friedman \cite{friedman2001greedy} who put forward that this technique can illustrate the impact of individual features on the predictions made by a machine learning model. Zhu et al. \cite{zhu2019forecasting} employed PDP to analyze how traditional ﬁnancing factors affect the credit risks of small and medium-sized enterprises (SMEs). Ozgur et al. \cite{ozgur2021machine} utilized PDP to prove that several bank-specific attributes and macroeconomic indicators play crucial roles in shaping the lending practices of banks. While PDP can visually represent the relationship between individual features and target variables, it might lack effectiveness when intricate relationships exist among features due to its inability to account for feature correlations \cite{molnar2020limitations}. ALE, an advanced version of PDP, is capable of capturing interactive effects among features. Liang et al. \cite{liang2022time} selected the ALE approach to explain the machine learning models of pricing European options. Christensen et al. \cite{christensen2021machine} adopted the ALE method to identify the key factors that have the most significant impacts on predicting volatility and examined the interplay effects among explanatory variables. However, ALE is theoretically not applicable to prediction functions that are non-differentiable, thereby limiting its applicability in the realm of interpretable machine learning \cite{molnar2020limitations}. By contrast, LIME is a local interpretability method that constructs a localized explanatory model around a specific instance to clarify the prediction of the original model for that instance \cite{ribeiro2016should}. As a result, LIME requires no considerations about the differentiability of the original model or any specific characteristics of its structure (i.e., model-agnostic). Park et al. \cite{park2021explainability} applied a LIME algorithm to calculate the feature importance based on tree-based models for predicting the likelihood of bankruptcy. Carta et al. \cite{carta2021explainable} focused on adopting LIME methods to aid in selecting relevant features for stock return prediction. LIME builds a local explanatory model, but its explanations can be influenced by randomness and selectivity. This is because the constructed local model is generated within the neighborhoods of the specific prediction instance in a random manner \cite{molnar2020limitations}. 

This article opts for SHapley Additive exPlanations (SHAP) as the method for explaining the self-attention GRU model. The concept of Shapley values traces its origins back to the work of Shapley \cite{shapley1953value} in 1953, which is used to investigate issues of fair allocation in game theory. Over time, the notion of Shapley values has gradually been introduced into the fields of data science and machine learning. In 2017, Lundberg and Lee \cite{lundberg2017unified} proposed the groundbreaking SHAP method, serving as a unified framework to interpret the outputs of any machine learning model. Specifically, the SHAP technique calculates the contribution of each feature to the final prediction results and these contribution values (i.e., Shapley values) are comparable at a global level. Furthermore, SHAP is model-agnostic and has the ability to capture the correlated contributions among features. The aforementioned advantages demonstrate that SHAP is a comprehensive interpretability tool kit distinct from PDP, ALE, and LIME. Jaeger et al. \cite{jaeger2021interpretable} employed Shapley values to compare the robustness of different asset portfolio strategies and deduce implicit decision rules. Bussmann et al. \cite{ bussmann2021explainable} applied Shapley values to assess the credit risks associated with borrowing through peer-to-peer lending platforms. Besides, there are other researchers such as Misheva et al. \cite{misheva2021explainable}, Gramegna et al. \cite{gramegna2021shap} and Moscato et al. \cite{moscato2021benchmark} who have contributed to the application of the SHAP method in the field of credit risk management. 
To the best of our knowledge, few articles have yet utilized the SHAP method for pricing American options, which lends significance to the study presented in this paper.

\section{Methodologies}
\subsection{Binomial tree method}
The binomial tree method, initially introduced by Cox, Ross, and Rubinstein \cite{cox1979option} in 1979, remains one of the most widely used approaches for pricing American options. Zhao \cite{zhao2018american} compared eight traditional American option pricing models and concludes that the classical binomial tree approach achieves the highest levels of accuracy and efficiency among all numerical methods. The binomial tree method has the benefits of a straightforward and easily understandable structure, user-friendly operations, and the ability to be personalized according to specific problems. 

\begin{figure}[h]
   \centering
   \includegraphics[width=6.5cm]{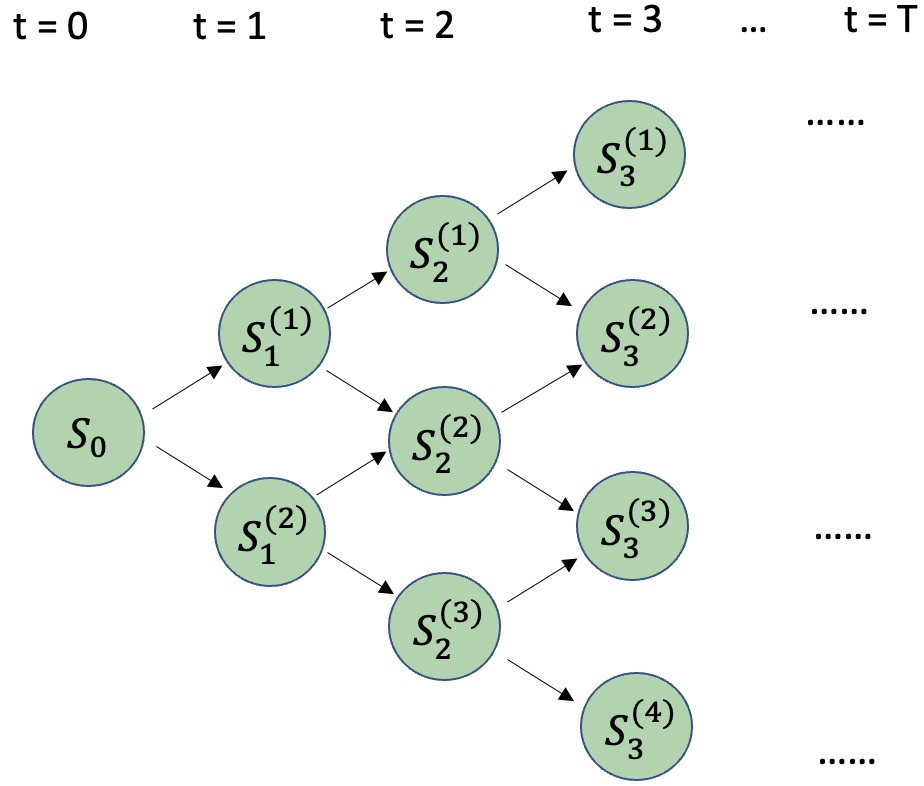}
    \caption{The underlying price tree}
    \label{The underlying price tree}
\end{figure}

An n-period binomial tree is a stochastic model that estimates the fluctuations of an underlying price as it evolves over a series of time intervals. To begin, the whole lifetime of the American option is divided into minuscule intervals denoted as $\Delta t$. In accordance with the principles of no-arbitrage, each node in the price tree experiences upward and downward movements. Over time, these movements aggregate to create an exponentially evolving trajectory of the price, effectively shaping a binomial tree representing the underlying price dynamics until the option expires. Subsequently, the terminal values of the option are calculated based on the option payoff function. Following this, any option value before the terminal node is determined as the higher value between its intrinsic value and its corresponding discounted expected payoff. To elaborate on the process further, the procedural steps are as follows:
\vspace{12pt}

\textbf{Step 1: Constructing the underlying price tree.} Every node of underlying prices $S_t$ at period $t$ has two trends: going up to $uS_t$  and going down to $dS_t$ , as indicated by equation \eqref{eq1}. Figure \ref{The underlying price tree} provides a simple illustration of an underlying price tree with three periods of nodes. As the number of periods increases, the quantity of nodes in the price tree correspondingly expands.

\begin{equation}
    S_{t+1}^{(i)} =
    \begin{cases}
        u\times S_t^{(i)}, & \text{if price increases from time } t \text{ to } t+1 \\
        d \times S_t^{(i)}, & \text{if price decreases from time } t \text{ to } t+1
    \end{cases}
    \label{eq1}
\end{equation}

Here, we require $u>e^{r\Delta t}>d>0$ where $r$ is the risk-free rate. The initial node $S_0$ in the underlying price tree is the spot price of the underlying asset. According to the Cox, Ross, and Rubinstein model \cite{cox1979option}, the upward factor $u$ and the downward factor $d$ can be calculated by using the volatility of underlying returns, and the time interval for each step is quantified in terms of years. Formulas \eqref{eq2} and \eqref{eq3} depict the multipliers for underlying price increases and decreases respectively.
\begin{equation}
    u = e^{\sigma \sqrt{\Delta t}}
    \label{eq2}
\end{equation}
\begin{equation}
    d = e^{-\sigma \sqrt{\Delta t}} = \frac{1}{u}
    \label{eq3}
\end{equation}
\vspace{1pt}

\textbf{Step 2: Determining the option value at the terminal node.} The value of an option is composed of intrinsic value and time value. After creating the underlying price map, the intrinsic value of each node can be calculated. The intrinsic value of an option refers to its payoff when exercised immediately. At the final period, options solely possess intrinsic value, as options are automatically exercised upon expiration if not exercised prematurely. As a result, we can obtain the option value of the terminal node at time T. Specifically, it can be shown as
\begin{equation}
    C_T^{(i)} =
    \begin{cases}
        \max(S_T - K, 0), & \text{for call options} \\
        \max(K - S_T, 0), & \text{for put options} 
    \end{cases}
    \label{eq4}
\end{equation}
where $S_T$ is the spot price of the underlying asset at the expiration date and $K$ is the strike price of the option.
\vspace{12pt}

\textbf{Step 3: Calculating option prices at previous nodes.} The biggest difference between American and European options lies in the fact that the American style allows for arbitrary exercise before the deadline, whereas the European style can only be exercised on the expiration date. At every node prior to the expiration date of an American option, calculations are made for both the immediate exercise payoff and the discounted expected payoff using the backward induction technique. The two values are then compared, and the higher of the two becomes the option price for that particular node. 
Under the context of risk-neutral equilibrium, the discounted expected value of the previous node is derived by weighting values of the subsequent two nodes based on their respective probabilities:
\begin{equation}
    C_{T-1}^{(i)} = [p C_T^{(i)} + (1-p) C_T^{(i+1)}]e^{-r\Delta t}
    \label{eq5}
\end{equation}
where the risk-free rate $r$ is constant in the basic setting. The factors $p$ and $1-p$ are the probabilities of underlying upward and downward movements respectively in a risk-neutral world. Cox et al. \cite{cox1979option} demonstrated that the binomial tree represents a special limiting case of the Black-Scholes model.  The risk-neutral probability $p$ is derived under the assumption of a geometric Brownian motion for the underlying asset, and it can be expressed as follows:
\begin{equation}
    p = \frac{e^{r\Delta t}-d}{u-d}
    \label{eq6}
\end{equation}

Since we have already obtained the option value of the terminal node from \textbf{step 2}, we can proceed to retroactively compute the discounted expected value at time $T-1$. Then, at each node, we compare its intrinsic value and its discounted expected value and the greater one becomes the ultimate option value. If the intrinsic value of one node is higher, it suggests that an immediate execution at that node is a more favorable strategy. By iteratively repeating this procedure, we can derive all option prices at previous nodes.

\subsection{Multilayer perception neural network}
In the realm of deep learning, the multilayer perceptron (MLP) stands as a commonly used and powerful model. The Multi-Layer Perceptron is an extension of the single-layer perception \cite{rosenblatt1957perceptron}, characterized by having multiple layers of neurons. The first layer of the MLP is typically referred to as the input layer, the intermediate layers are known as hidden layers, and the final layer is the output layer. The number of hidden layers in an MLP is not fixed, allowing for flexibility in choosing the appropriate number based on practical processing needs. Similarly, there are no constraints on the number of neurons in the input layer and output layer. MLPs find extensive application in tasks such as classification and regression.

At the core of the MLP are neurons (or nodes), which simulate connections between human neurons. Each neuron has an activation function, responsible for processing input signals and generating outputs.  In an MLP, each neuron is connected to all neurons in the previous layer, and each connection has a weight, determining the significance of the signal.

The input layer receives raw data and passes it to the hidden layers, which progressively extract features and transform them into higher-level representations. Each layer can be represented by the following formula:

\begin{equation}
a^{(l)} = f(W^{(l)}a^{(l-1)} + b^{(l)})
\label{eq7}
\end{equation}

Here, \(l\) denotes the index of the current layer, \(a^{(l)}\) is the output of that layer, \(f\) is the activation function, and \(W^{(l)}\) and \(b^{(l)}\) are the weight matrix and bias vector, respectively. Positive weights amplify the transmission of input signals among neurons, while negative weights are commonly employed to dampen signal propagation between neurons. 
Positive and negative weights, along with biases, are typically learned automatically by neural network algorithms through backpropagation during the training process, enabling the network to adapt to the given task. 
The output of the final hidden layer is passed to the output layer, generating the ultimate prediction.
We can further illustrate the feedforward and backpropagation mechanisms of MLP through the following diagram:

\begin{figure}[h]
   \centering
   \includegraphics[width=\linewidth]{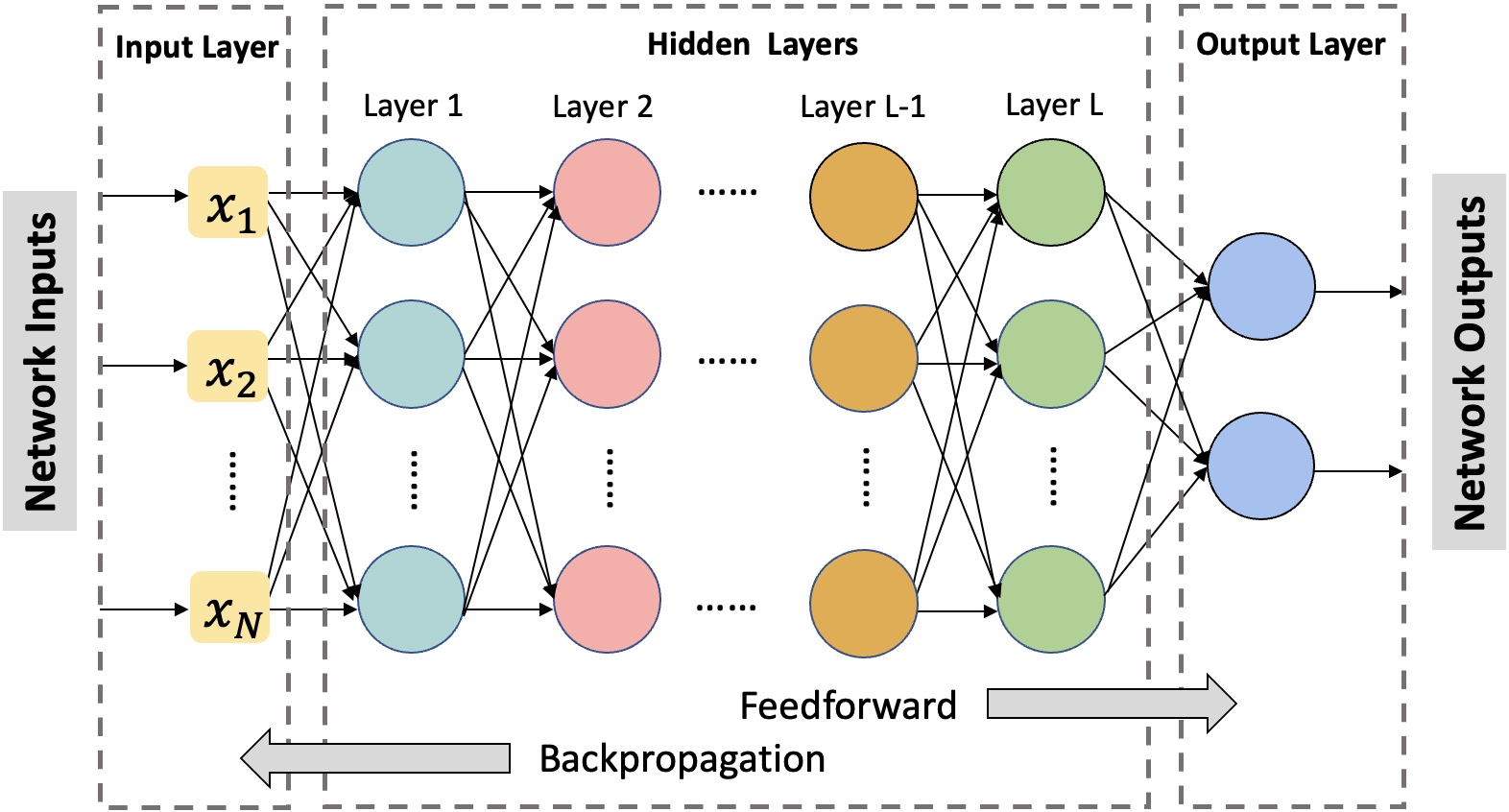}
    \caption{Multilayer perception neural network}
    \label{MLP}
\end{figure}
In the feedforward phase of the MLP, input data traverses from the input layer through hidden layers to the output layer, generating predictive results. Each neuron computes and transmits signals utilizing weights, biases, and activation functions. This process progressively transforms data, extracting features and producing final outputs. The feedforward process does not involve adjustments of weights and biases. In the subsequent backpropagation phase, the difference between predicted outcomes and actual targets is quantified by computing the loss function. Starting from the output layer, gradients of the loss with respect to each weight and bias are calculated through the chain rule. 
Afterwards, we need to choose appropriate optimizer algorithms to update each weight and bias iteratively.

Adaptive moment estimation (Adam) \cite{kingma2014adam} is a popular optimization algorithm used in training machine learning models,
especially deep neural networks. Adam offers an efficient and adaptive optimization approach compared to the traditional stochastic gradient descent (SGD). The main idea behind the Adam algorithm is to adaptively adjust the learning rates for different parameters in the model. It maintains an exponentially decaying average of past gradients and their
squared values. 
Additionally, it also keeps track of the momentum-like moving average of gradients. The Adam update rule involves computing moving averages for both the first moments (mean) and second moments (uncentered variance) of gradients. 
These moments are then used to update the model parameters. The algorithm also includes bias correction terms to address the initialization bias that occurs in the early iterations. To be more specific, the whole computation procedure can be shown as follows:

Given the loss function $L(\theta)$, where $\theta$ represents the model parameters (weights and biases), the goal is to find the values of $\theta$ that minimize the loss.

\textbf{Step 1:} Initialize the parameters $\theta$ as $\theta_0$.

\textbf{Step 2:} Choose hyperparameters including learning rate $\eta$, exponential decay rates for moment estimates $\beta_1, \beta_2 \in [0, 1) $, and smoothing term $\epsilon$ to prevent division by zero.

The gradient of the loss with respect to the parameters $\theta$ is computed as:
\begin{equation}
\nabla L(\theta) = \left( \frac{\partial L(\theta)}{\partial \theta_1}, \frac{\partial L(\theta)}{\partial \theta_2}, \ldots, \frac{\partial L(\theta)}{\partial \theta_n} \right)
\label{eq8}
\end{equation}
where $n$ is the number of parameters.

\textbf{Step 3:} Initialize moment estimates:
\begin{equation}
m_0 = 0
\end{equation}
\begin{equation}
v_0 = 0
\end{equation}

\textbf{Step 4:} For iteration $t = 1, 2, \ldots$:
\begin{equation}
\begin{aligned}
m_t &= \beta_1 \cdot m_{t-1} + (1 - \beta_1) \cdot \nabla L(\theta_t)
\end{aligned}
\end{equation}
\begin{equation}
\begin{aligned}
v_t &= \beta_2 \cdot v_{t-1} + (1 - \beta_2) \cdot (\nabla L(\theta_t))^2
\end{aligned}
\end{equation}
\begin{equation}
\begin{aligned}
\hat{m}_t &= \frac{m_t}{1 - \beta_1^t}
\end{aligned}
\end{equation}
\begin{equation}
\begin{aligned}
\hat{v}_t &= \frac{v_t}{1 - \beta_2^t}
\end{aligned}
\end{equation}
\begin{equation}
\begin{aligned}
\theta_{t+1} &= \theta_t - \eta \cdot \frac{\hat{m}_t}{\sqrt{\hat{v}_t} + \epsilon}
\end{aligned}
\end{equation}
Here, $\nabla L(\theta_t)$ is the gradient at time step $t$, $m_t$ and $v_t$ represent the first and second moment estimates at time step $t$, $\hat{m}_t$ and $\hat{v}_t$ are bias-corrected moment estimates,
$\beta_1^t$ and $\beta_2^t$ denote $\beta_1$ and $\beta_2$ to the power $t$.

Repeat \textbf{step 4} for a fixed number of iterations or until convergence. The process aims to iteratively update the parameters using the moment estimates to adaptively adjust the learning rates for each parameter, resulting in efficient optimization. 
The MLP neural network iterates through the sequence of feedforward and backpropagation processes, continuously updating parameters until the model reaches a sufficient number of training iterations or the intended loss objective.

It is worth noting that the choice of hyperparameters for an MLP neural network can significantly impact the training outcomes of the model.\footnote{This holds true for other deep learning models as well.} The common hyperparameter types are outlined below:
\begin{itemize}
\item \textbf{Number and Size of Hidden Layers:} Determine the depth and width of MLP. More hidden layers and neurons can provide stronger feature learning capabilities, but may also increase the risk of overfitting.
\item \textbf{Activation Functions:} Introduce nonlinearity to the neural networks. Common activation functions include ReLU, Sigmoid, and Tanh.
\item \textbf{Learning Rate:} Determines the step size for each parameter update. A too-small learning rate may lead to slow convergence, while a too-large learning rate might cause oscillation or unstable convergence.
\item \textbf{Batch Size:} Specifies the number of training samples used to calculate gradients in each update. Larger batch sizes can speed up training but might also increase the memory requirements of computers.
\item \textbf{Number of Epochs:} Defines how many times the training data is iterated through the model. Too many epochs can lead to overfitting, while too few might prevent the model from fully learning.
\item \textbf{Optimizer:} Algorithms used to update weights, such as Adam, SGD, RMSProp, AdaGrad, and so on.
\item \textbf{Regularization:} Techniques utilized to reduce model overfitting by adding penalty terms to the loss function, thereby constraining the complexity of parameters and enhancing generalization capabilities of the model. Common regularization approaches encompass L1 regularization, L2 regularization, Dropout, and Early Stopping.
\end{itemize}
Tuning these hyperparameters can significantly impact model performance and often requires finding the best combinations through methods like cross-validation.

In summary, the multilayer perceptron (MLP), as a type of multi-layer neural network model, has demonstrated its powerful capabilities in the field of machine learning. Through layered neural connections and appropriately chosen activation functions, the MLP is able to capture intricate data patterns and nonlinear relationships, making it suitable for various tasks. Its training approach involves defining network architecture, selecting suitable loss functions, and utilizing optimization algorithms for parameter updates. The flexibility and expressive power of MLP make it an indispensable tool to provide robust solutions for data-driven problems.

\subsection{Long short-term memory neural network}
Long short-term memory (LSTM) \cite{hochreiter1997long} is designed to overcome the vanishing gradient problem in traditional recurrent neural networks (RNNs). LSTM networks are particularly effective in capturing long-range dependencies and patterns in sequential data. The vanishing gradient problem arises when gradients become too small as they propagate backward through time, causing difficulties in learning and retaining information over long sequences. LSTM addresses this challenge by introducing specialized mechanisms to control the flow of information and gradients within the network.

\begin{figure}[h]
    \centering
    \includegraphics[width=\linewidth]{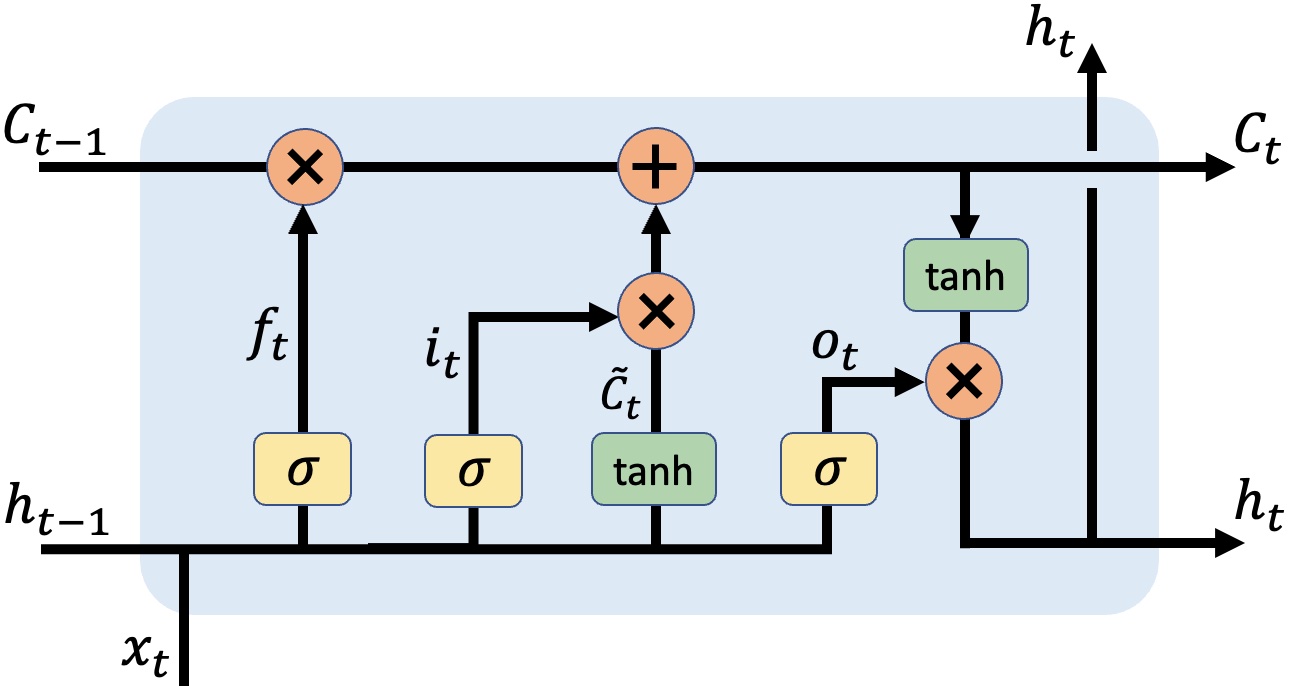}
    \caption{LSTM cell}
    \label{LSTM}
\end{figure}

Figure \ref{LSTM} illustrates the basic architecture of an LSTM cell. An LSTM cell consists of three main components:
\begin{itemize}
\item \textbf{Cell State ($C_t$)}: The cell state serves as a memory that can store and carry information across different time steps. It allows LSTM to remember or forget information as needed.

\item \textbf{Hidden State (\(h_t\))}: The hidden state captures the relevant information from the current input and the previous hidden state. It acts as the output of the LSTM cell and is also used in the next time step.

\item \textbf{Gates}: LSTM uses three gates to control the flow of information: the forget gate (\(f_t\)), the input gate (\(i_t\)), and the output gate (\(o_t\)). These gates regulate the information that is either discarded or passed through the cell state and hidden state. 
\end{itemize}

The detailed calculations within an LSTM cell are described by the following equations:
\begin{align}
    f_t & = \sigma(W_f \cdot [h_{t-1}, x_t] + b_f) \\
    i_t & = \sigma(W_i \cdot [h_{t-1}, x_t] + b_i) \\
    \tilde{C}_t & = \tanh(W_C \cdot [h_{t-1}, x_t] + b_C) \\
    C_t & = f_t \odot C_{t-1} + i_t \odot \tilde{C}_t \\
    o_t & = \sigma(W_o \cdot [h_{t-1}, x_t] + b_o) \\
    h_t & = o_t \odot \tanh(C_t)
\end{align}
Where:
\begin{itemize}
    \item \(x_t\) is the input at time step \(t\).
    \item \(h_{t-1}\) is the hidden state from the previous time step.
    \item \(W_f, W_i, W_C, W_o\) are weight matrices for the gates.
    \item \(b_f, b_i, b_C, b_o\) are bias vectors for the gates.
    \item \(\sigma\) is the sigmoid activation function.\footnote{The expression of the sigmoid equation is: $\sigma(x) = \frac{1}{1 + e^{-x}}$. The main advantage of the sigmoid function is that it restricts its output within the range of 0 to 1, making it suitable for probability representation and activation functions. }
    \item \(\odot\) denotes element-wise multiplication.
\end{itemize}   
The forget gate $f_t$ assesses which information to forget from the cell state, incorporating prior hidden state $h_{t-1}$ and current input $x_t$. This yields a weight $f_t$ that controls the degree of forgetting. Subsequently, the input gate $i_t$ computes how much new data to include in the cell state, relying on prior hidden state $h_{t-1}$ and current input $x_t$. A candidate cell state $\tilde{C}_t$ forms from the blend of previous hidden state $h_{t-1}$ and current input $x_t$, refined through a tanh activation. The cell state $C_t$ updates collaboratively, melding insights from $f_t$, $i_t$, and $\tilde{C}_t$. The output gate $o_t$ governs data flow to the next hidden state and output, influenced by past hidden state $h_{t-1}$ and current input $x_t$. Ultimately, hidden state $h_t$ emerges, fusing information from $C_t$ via output gate $o_t$, which then propels insights to the next time step and model output. These gate interactions empower LSTMs to adeptly manage sequences, capturing intricate dependencies and adapting seamlessly.

\subsection{Gated recurrent unit neural network}
Similar to LSTMs, the gated recurrent unit (GRU) \cite{cho2014learning} is another RNN architecture crafted to combat the vanishing gradient problem. It also excels at capturing long-range dependencies and patterns within sequential data.
However, GRUs offer a streamlined architecture with fewer parameters, which can lead to faster training and optimization. GRUs also combine the cell state and hidden state into a single "hidden state," simplifying memory management compared to LSTMs. Moreover, GRUs employ only two gates, the reset gate and the update gate, in contrast to the three gates of LSTMs, which can result in simpler learning and reduced overfitting. These advantages make GRUs a compelling choice, especially when computational resources are limited or when dealing with datasets of moderate sequence lengths.

\begin{figure}[h]
    \centering
    \includegraphics[width=7.5cm]{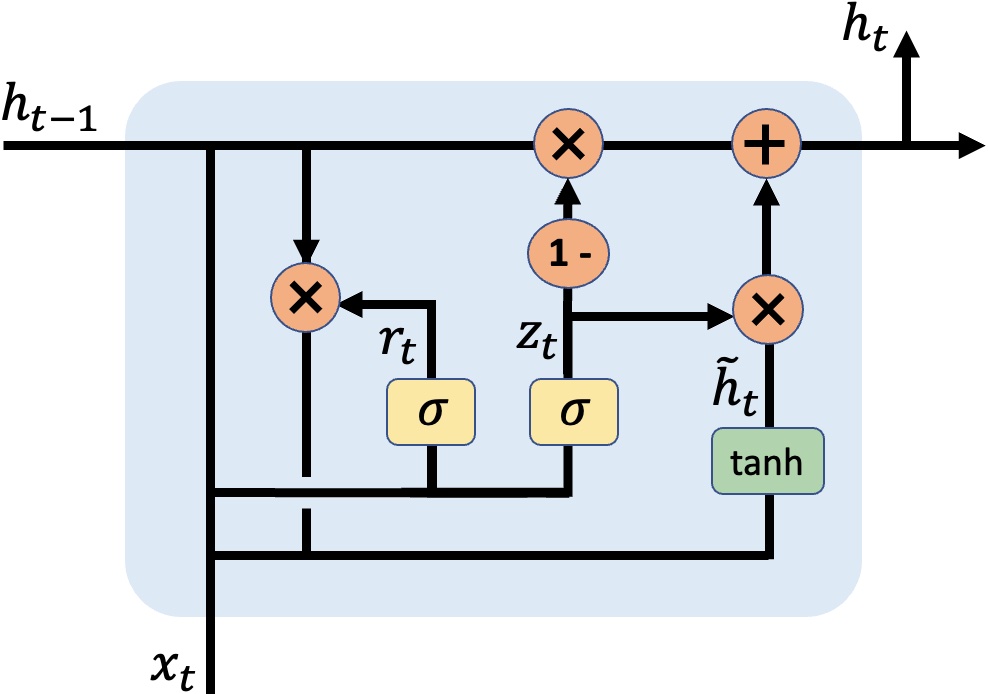}
    \caption{GRU cell}
    \label{GRU}
\end{figure}

Figure \ref{GRU} portrays the fundamental structure of a GRU cell. A GRU cell comprises two parts:
\begin{itemize}
\item \textbf{Hidden State ($h_t$)}: The hidden state acts as both memory and carrier of information across various time steps. This duality empowers the GRU to dynamically decide whether to preserve or discard information based on the context.

\item \textbf{Gates}: There are two gates in a GRU cell: the reset gate $r_t$ decides when to omit prior hidden state data, while the update gate $z_t$ governs the balance between newly presented and preceding information. These gates allow GRU to adjust the information processing dynamically, making it more flexible in handling sequence data.
\end{itemize}

The following mathematical formulas elucidate how GRU propagates and updates information:
\begin{align}
    z_t & = \sigma(W_z \cdot [h_{t-1}, x_t]) \\
    r_t & = \sigma(W_r \cdot [h_{t-1}, x_t]) \\
    \tilde{h}_t & = \tanh(W_h \cdot [r_t \odot h_{t-1}, x_t]) \\
    h_t & = (1 - z_t) \odot h_{t-1} + z_t \odot \tilde{h}_t
\end{align}
Where:
\begin{itemize}
    \item $x_t$ represents the input at time step $t$.
    \item $h_{t-1}$ is the hidden state from the previous time step.
    \item $W_z, W_r, W_h$ are weight matrices associated with the gates.
    \item $\sigma$ denotes the sigmoid activation function.
    \item $\odot$ signifies element-wise multiplication.
\end{itemize}   
Specifically, the update gate $z_t$ determines how much information to retain from the previous hidden state $h_{t-1}$ and the current input $x_t$. The reset gate $r_t$ controls whether to reset the hidden state to adapt to new input. A candidate hidden state $\tilde{h}_t$ is created by combining the previous hidden state $h_{t-1}$, which is weighted by the reset gate $r_t$, with the current input $x_t$. This combined representation is then refined using a tanh activation function. $\tilde{h}_t$ signifies potential update information for the current time step. It is utilized to decide whether to refresh the current hidden state, thereby regulating information flow and memory update. 
The hidden state $h_t$ is further fine-tuned by fusing it with the previous hidden state $h_{t-1}$ and the candidate hidden state $\tilde{h}_t$. This fusion process is controlled by the update gate $z_t$, facilitating the transfer of information to the following time step and contributing to the model outputs.

\subsection{Self-attention mechanism}
While RNN models like LSTM and GRU have advantages in handling time series data, if the time series becomes rather long, the memory burden of these models gradually increases, making them susceptible to the issue known as "long-term dependency". This implies that they struggle to effectively retain and utilize information from earlier time steps. In such cases, these models might "forget" crucial information from earlier time steps, consequently affecting their comprehension and prediction of the entire sequence. However, the emergence of the attention mechanism provides a good solution to this problem. The attention mechanism, first proposed by Bahdanau et al. \cite{bahdanau2014neural}, allows models to focus on key information from different parts of the sequence during predictions, regardless of the sequence length. Attention mechanisms come in various forms, each designed to enhance the ability of models to focus on relevant information within a given context. One notable type is self-attention \cite{vaswani2017attention}, which has garnered considerable notice due to its effectiveness in capturing dependencies and patterns within a sequence.

The self-attention mechanism allows the model to determine the importance of elements within the same sequence based on their interrelationships. By comparing the relevance of each element with others, the model can identify which elements are more significant in a given context, thus assigning varying weights to different elements.
This is accomplished by forming three sets of projections, known as the Query ($Q$), Key ($K$), and Value ($V$) matrices. The Query projections capture the information a model is looking for, the Key projections encode the importance of elements, and the Value projections hold the actual values to be combined based on the attention scores.

For example, given an input sequence represented by the matrix $F \in \mathbb{R}^{T \times d_h}$, the self-attention mechanism is defined as follows:
\begin{align}
Q & = \text{query\_layer}(F) \\
K & = \text{key\_layer}(F) \\
V & = \text{value\_layer}(F) \\
S & = \frac{Q \cdot K^\intercal}{\sqrt{d_k}} \\
W & = \text{softmax}(S) \\
O & = W \cdot V
\end{align}

\begin{figure}[h]
  \centering
  \includegraphics[width=\linewidth]{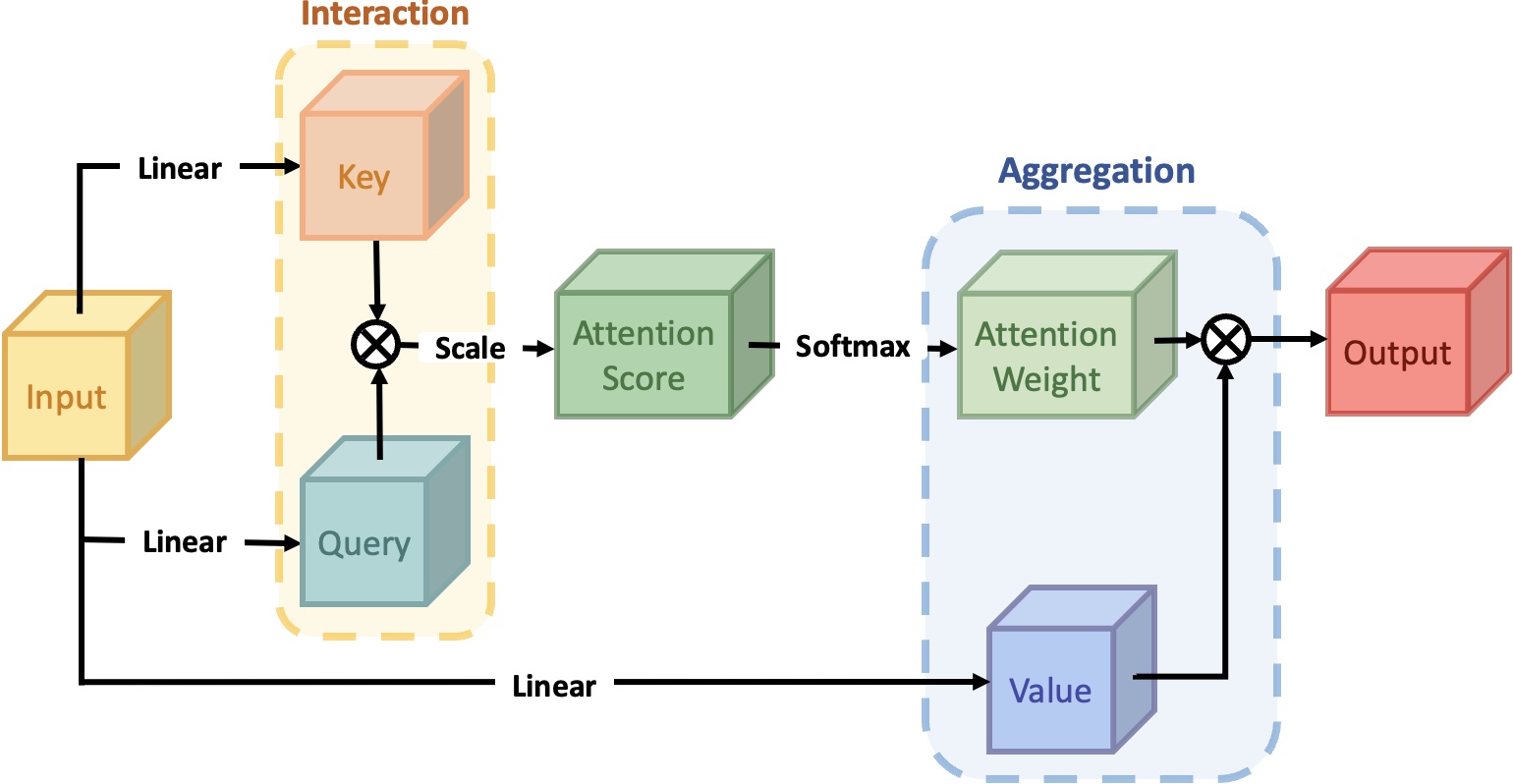}
  \caption{Self-attention mechanism}
  \label{Self-attention Mechanism}
\end{figure}

Where:
\begin{itemize}
  \item $\text{query\_layer}$, $\text{key\_layer}$, $\text{value\_layer}$ are linear layers projecting the input $F$ into the Query, Key, and Value spaces respectively.
  \item $T$ is the sequence length, and $d_h$ is the hidden dimension of the input.
  \item $S$ is the vector of attention scores, and $d_k$ denotes the dimension of $K$.
  \item The softmax function is expressed as $\text{softmax}(\mathbf{x}_i) = \frac{e^{x_i}}{\sum_{j=1}^n e^{x_j}}$, where $\mathbf{x}_i$ is the $i$-th element of the input vector.
  \item $W$ is the calculated attention weight, and $O$ represents the final self-attention output.
\end{itemize}
First, the self-attention mechanism computes $Q$, $K$, and $V$ matrices respectively. 
Then, the dot product of the $Q$ and $K$ matrices is scaled by $\frac{1}{\sqrt{d_k}}$ to obtain the attention scores. The softmax function normalizes these scores to produce weights that indicate the importance of different elements.
Finally, the weighted sum of Value vectors is computed to generate the attention output $O$.
The entire computational process is directly illustrated by figure \ref{Self-attention Mechanism}. 


The self-attention mechanism can be visually understood as a process of aligning each element with the others in the sequence and assigning weights to them based on their relevance to the current context. This allows the model to capture both local and global dependencies, facilitating the incorporation of long-range information.
The ability of self-attention to grasp relationships across varying distances makes it exceptionally well-suited for tasks involving sequential data, such as natural language processing, audio signal processing, and time series forecasting. It enables models to capture intricate patterns and relationships that are crucial for accurate predictions.
In the context of LSTM and GRU architectures, integrating self-attention can enhance their capacity to capture complex temporal dependencies and result in improved performance on tasks requiring sequence modeling.

\subsection{The interpretable method - Shapley Value}
SHapley Additive exPlanations (SHAP) is a potent and effective tool for explaining machine learning models. It can be applied to various types of models and diverse prediction tasks, aiding in the comprehension of the impact of each feature on predictions globally. 

SHAP originates from the concept of Shapley values from cooperative game theory, which was proposed by Shapley \cite{shapley1953value}. 
The Shapley value of a player in a cooperative game is defined as the average contribution of that player across all possible coalitions. Mathematically, the Shapley value of player \(i\) is given by:
\vspace{2pt}  
\begin{equation}
    \phi_i(v) = \sum_{S \subseteq N \setminus \{i\}} \frac{|S|!(|N|-|S|-1)!}{|N|!} \left[ v(S \cup \{i\}) - v(S) \right]
\end{equation}
Where:
\begin{itemize}
    \item \(N\) is the set of all players and \(|N|\) is the number of all players.
    \item $S$ is the coalition without player $i$ and \(|S|\) denotes the number of players in \(S\).
    \item \(v(S)\) represents the value of coalition \(S\).
    \item \(v(S \cup \{i\})\) represents the value of coalition \(S\) with player \(i\).
    \item \(\sum_{S \subseteq N \setminus \{i\}}\) indicates a summation over all sets \(S\) that are subsets of \(N\) excluding player \(i\).
\end{itemize}
$\phi_i(v)$ is the contribution value function for player $i$. It computes the weighted average sum of the value contribution resulting from each change in various possible coalitions $i$ can join. 
$\frac{|S|!(|N|-|S|-1)!}{|N|!}$ is a coefficient used to adjust the influence of coalition size on the contribution. It is a binomial coefficient that takes into account the coalition size to ensure fairness in the calculations.
$v(S \cup \{i\}) - v(S)$ calculates the difference between the value of the coalition when player $i$ is present and when player $i$ is absent. This part displays the additional value brought by player $i$ joining coalition $S$.

The traditional Shapley method involves considering various possibilities of coalitions $S$ that do not include player $i$, and then obtaining the weighted average of the marginal contributions brought by including player $i$ in $S$. This yields the average contribution of player $i$ among all players. The theoretical logic above can be extended to explain machine learning models by treating the features of the model as players. In 2017, Lundberg et al. \cite{lundberg2017unified} introduced SHapley Additive exPlanations (SHAP) based on the traditional Shapley value concept to explain machine learning models and released the corresponding codes. Since then, SHAP has gained widespread adoption across a wide spectrum of domains, ranging from natural language processing to healthcare and finance.
In specific, SHAP quantifies the average contribution of each feature to a model's final prediction, providing interpretability to the prediction results. 
The SHAP value for feature \(x_i\) with respect to instance \(x\) can be expressed as:
\begin{equation}
    \phi_i(x) = \sum_{S \subseteq N \setminus \{i\}} \frac{|S|!(|N|-|S|-1)!}{|N|!} \left[ f(x_S \cup \{i\}) - f(x_S) \right]
\end{equation}
Where:
\begin{itemize}
    \item $N$ is the set of all features and \(|N|\) is the number of all features.
    \item $|S|$ indicates the number of features in subset $S$.
    \item $f(x)$ represents the model's prediction for instance $x$.
    \item $x_S$ denotes the subset of features in instance $x$ indexed by $S$. 
\end{itemize}
This formula iterates over all possible feature subsets $S$ excluding feature $i$, calculating the difference in predictions when feature $i$ is included compared to its absence. The SHAP value is calculated as a weighted sum of these differences, with weights determined by the binomial coefficients. The resulting SHAP values provide a comprehensive understanding of how each feature of a given instance influences the ultimate prediction, aiding model interpretability.

Unlike traditional interpretability techniques, SHAP captures the nuanced interactions between features and their combined contributions to prediction outcomes, offering a more comprehensive and insightful perspective.
Besides,  SHAP is able to assign a consistent and fair distribution of feature importance across different samples. This attribute ensures that the interpretability insights remain robust and reliable even when applied to various instances. 

As machine learning models continue to evolve and become more complex, the need for interpretable explanations becomes increasingly crucial. SHAP's adaptability, mathematical rigor, and capability to elucidate intricate relationships make it an indispensable tool for both practitioners and researchers seeking to uncover the underlying mechanisms of their models. By contributing to a deeper understanding of model behavior, SHAP paves the way for informed decision-making and responsible AI deployment.

\section{Data preparation}
\subsection{Introduciton of data}
The data for our empirical experiments are daily closing prices of SPY and SPY call option for the 3-year period from January 2020 to December 2022.\footnote{The data is sourced from Kaggle: https://www.kaggle.com/datasets/kylegraupe/spy-daily-eod-options-quotes-2020-2022} 

SPY options are a type of financial derivative that allows investors to trade the price movement of the SPDR S$\&$P 500 ETF (Exchange-Traded Fund), commonly known as SPY. SPY is one of the most popular ETFs and tracks the performance of the S$\&$P 500 index, which consists of 500 large-cap U.S. companies.

Options are contracts that give investors the right, but not the obligation, to buy (call option) or sell (put option) an underlying asset, such as stocks or ETFs, at a predetermined price (strike price) within a specified period (expiration date). SPY options are particularly attractive to investors seeking exposure to the broader stock market without directly trading individual stocks. In our study, we focus on the call style of SPY options, which investors employ in pursuit of profits through potential increases in SPY prices.
All equity (single-stock) and ETF options are American options, allowing them to be exercised at any point before or on the expiration date. Therefore, the primary research topic of this paper revolves around conducting a comprehensive analysis of American option pricing. American options represent the prevalent actively traded options in the market, characterized by a heightened degree of complexity and flexibility in pricing.

SPY options are traded on major options exchanges and have standardized contract sizes, expiration dates, and strike prices. Due to their popularity and importance as a market benchmark, the trading volume of SPY options is relatively larger compared to options of individual stocks. The dataset used in our study consists of 2,794,966 records, and the substantial amount of data is advantageous for enabling machines to "learn" and generate more stable models while reducing the risk of overfitting.

\subsection{Data preprocessing}
This paper presents a comparison between the classical binomial tree models and various deep learning models in the pricing of SPY call options. To ensure comparability, all features are kept consistent across different models. Similar to traditional pricing models, our features include underlying spot price, strike price, days to maturity\footnote{"days to maturity" refers to the number of days remaining from the current trading day to the expiration date.}, underlying volatility, and risk-free interest rate. Furthermore, we include the moneyness of options, which is the ratio of spot price to strike, as an additional feature during the training of deep learning models. Moneyness of options is a crucial concept in options trading and valuation. It signifies the relationship between the current spot price of the underlying asset (such as a stock or index) and the strike price of the option. Moneyness helps to categorize options into three different classifications based on their intrinsic value and potential for profitability:
\begin{itemize}
    \item \textbf{Out-of-the-Money (OTM) Options:} Options are labeled "out-of-the-money" when the strike price is less favorable compared to the current spot price. Call options are OTM if the spot price is below the strike price or if moneyness is less than 1. OTM options generally have no intrinsic value, and their value is mostly derived from time value and market expectations.
    \item \textbf{At-the-Money (ATM) Options:} An option is "at-the-money" when the spot price is approximately equal to the strike price or when moneyness is close to 1. ATM options often have higher time value and are subject to price changes based on market movements.
    \item \textbf{In-the-Money (ITM) Options:} An option is considered "in-the-money" when its strike price is favorable compared to the current spot price of the underlying asset. For call options, this occurs when the spot price is higher than the strike price or when moneyness is bigger than 1. ITM options tend to have intrinsic value and are more likely to be exercised.
\end{itemize}
Integrating moneyness into the deep learning model enhances its ability to uncover the nuanced impact of the relative correlation between spot price and strike on the final prediction (i.e., option price), as compared to considering spot price and strike individually. Particularly, within our dataset, options are classified as follows based on moneyness:
\begin{itemize}
    \item \textbf{OTM option:} When $\text{moneyness} \leq 0.97$
    \item \textbf{ATM option:} When $0.97 < \text{moneyness} \leq 1.03$
    \item \textbf{ITM option:} When $\text{moneyness} > 1.03$
\end{itemize}

Apart from the moneyness-based data classification criterion, we further employ the days to maturity as an additional classification criterion. This means that within the OTM, ATM, and ITM categories of options, we subdivide them into five subcategories based on days to maturity: 
\begin{itemize}
    \item days to maturity $\leq 9$ days
    \item $9$ days $<$ days to maturity $\leq 30$ days
    \item $30$ days $<$ days to maturity $\leq 90$ days
    \item $90$ days $<$ days to maturity $\leq 180$ days
    \item days to maturity $> 180$ days
\end{itemize}
As a result, the data on the 3-year period SPY call options can be partitioned as 15 non-overlapping datasets based on the criteria of moneyness and days to maturity. Our intention in partitioning the data according to days to maturity is to align different days to maturity values with their respective risk-free interest rates and volatilities of underlying returns. This approach enables a better estimation of interest rates and volatilities, thereby facilitating the analysis of their impacts on model predictions. We utilize treasury yields of various maturities as estimates for risk-free interest rates.\footnote{The treasury yields are obtained from the US Treasury Resource Center: https://home.treasury.gov/policy-issues/financing-the-government/interest-rate-statistics?data=yield} 

The selection of volatility has consistently remained a crucial topic in the field of option pricing. In addition to treating volatility as a constant in traditional models such as Black-Scholes model, there have been two common methods in previous literature. One approach directly uses historical realized volatility as a predictive value \cite{kelly1994valuing, hutchinson1994nonparametric}, while the other entails forecasting volatility using various prediction techniques such as implied volatility \cite{malliaris1993neural,anders1998improving}, statistical methods \cite{meissner2001capturing,pande2006new}, and neural networks \cite{malliaris1996using,amornwattana2007hybrid}. Even, Yao et al. \cite{yao2000option} did not treat volatility as a feature of the neural network model, considering that the model should be able to automatically capture information on volatility since it already incorporates the underlying spot prices. 

This paper employs the implied volatility approach but the novelty is that we align implied volatility indices with their respective days to maturity. Specifically, we select CBOE Short-Term Volatility Index (VXST), CBOE Volatility Index (VIX), CBOE 3-Month Volatility Index (VIX3M), CBOE Mid-Term Volatility Index (VXMT), and CBOE One-Year Volatility Index (VIX1Y) to approximate the SPY's volatility corresponding to five types of days to maturity.\footnote{The data of all these implied volatility indices are downloaded from Yahoo Finance: https://finance.yahoo.com/} All these indices are constructed and maintained by the Chicago Board Options Exchange (CBOE), serving as indicators to gauge market anticipated volatility. The underlying asset of the five indices is S$\&$P 500 index, which makes them suitable for estimating the volatility of SPY.
In a more detailed manner, VXST measures short-term market volatility, typically around 9 days, so we use it as a representation of the "volatility" feature in model building when options' days to maturity are no more than 9 days. VIX quantifies market volatility expectations over the next 30 days, which we use as "volatility" when $9$ days $<$ days to maturity $\leq 30$ days. VIX3M extends the volatility prediction horizon to three months and we employ it if $30$ days $<$ days to maturity $\leq 90$ days. VXMT focuses on a medium-term outlook, capturing the market's expectations for volatility over a six-month time frame, thus it is appropriate for estimates of "volatility" when $90$ days $<$ days to maturity $\leq 180$ days. VIX1Y provides a longer-term perspective on market volatility expectations over a one-day period, which we use to represent the "volatility" when days to maturity $\geq 180$ days. Implied volatility is a volatility level deduced from option market prices, providing a more accurate reflection of market participants' anticipations for future volatility compared to historical realized volatility. In selecting implied volatility, we account for different remaining days to maturity, allowing for a more refined estimation of the "volatility" input for all models.

In summary, when we do not take historical data of options into account, our models encompass a set of six features as inputs: spot price, strike, moneyness, days to maturity, volatility, and interest rate.
As for the target output of pricing models, we choose the mid-price of the options, which is the average value of the bid price and ask price recorded in option transactions.

\section{Experimental results}
\subsection{Model pricing process}
\subsubsection{Implementation details} 
When pricing SPY call options using the binomial tree model, unlike in conventional practices where risk-free interest rates and underlying volatility remain constant, we adopt values that vary with the corresponding expiration date.
Within the underlying price tree map, the initial node is set by the underlying spot price, and we assume the time interval for each step $\Delta t$ is equal to $\frac{1}{252}$.\footnote{Assuming there are 252 trading days in a year.} Then we can get the whole price tree and the option price map can be derived with the backward induction technique. The first node of the option price map is our predicted SPY call value from the binomial tree approach. 

We conduct feature normalization in advance for all deep learning machine models, ensuring uniform scales across all features. This prevents features of different magnitudes from affecting the stability of model training and even potentially causing convergence issues. We utilize Max-Min normalization for all features. Assuming $x$ denotes the individual feature and the normalization equation is shown as:
\begin{equation}
   x_{\text{normalized}} = \frac{x - x_{\text{min}}}{x_{\text{max}} - x_{\text{min}}} 
   \label{eq28}
\end{equation}
Through equation \eqref{eq28}, all features can be scaled to numbers between 0 and 1. After normalization, we split each of the 15 datasets into training, validation, and test sets using proportions of 70\%, 10\%, and 20\%, respectively. It is necessary to note that we shuffle the samples before partitioning to alleviate the potential influence of the original data's chronological order on model performance evaluation. Shuffling the data helps in assessing the generalization performance of models more accurately. 
For a deep learning model, performing well on the training set does not guarantee equivalent performance on the test set.\footnote{It is called overfitting when a machine learning model performs well on the training set but poorly on out-of-sample data. The introduction of a validation set can help mitigate this potential risk.} 
Therefore, we need to set aside a validation set to help us understand the model's performance on out-of-sample data once the model has been trained. In other words, we train the model and then assess its performance on the validation set for multiple epochs. The final "best" model we decide to save is the one that performs best on the validation set. 
Subsequently, we evaluate the best model's performance on the test set to evaluate its out-of-sample effectiveness. This procedure ensures that the error computed on the test set better reflects the model's performance in real-world scenarios.

We experiment with four types of deep learning models: MLP, LSTM, self-attention LSTM, and self-attention GRU. The specific MLP architecture includes one input layer, two hidden layers with 64 neurons in each layer, and one output layer. 
The activation function between each layer of MLP is Rectified Linear Unit (ReLU). The ReLU function is defined as $f(x) = \max(0, x)$, which is similar to the payoff function of options. 
ReLU is one of the key factors that allow machine learning models to learn non-linear relationships. 
The learning rate, batch size, and number of epochs for the MLP are set to 0.0001, 1024, and 2000, respectively.\footnote{The batch size for all deep learning models is set to 1024, and this will not be reiterated later.} Besides, we select the Mean Squared Error (MSE) loss function and Adam optimizer for all models. 

Regarding the LSTM model, we conduct empirical research first on samples without historical data and then on time series samples considering historical data. 
The LSTM model is constructed with six LSTM layers and one fully connected layer\footnote{The fully connected layer essentially involves a linear transformation from one feature space to another.} before the final output. The learning rate and number of epochs for the LSTM model without historical data are set to 0.001 and 2000. We also attempt using the same learning rate as the MLP model, but the performance was not satisfactory. An improvement in LSTM performance can be observed after increasing the learning rate to 0.001\footnote{Here after, the learning rate for the rest models is set to 0.001, and this will not be reiterated later.}. This is because, as neural network models become more complex, a larger learning rate can expedite the model's movement in parameter space, leading to quicker convergence towards a relatively favorable solution of the loss function.

The ability of LSTM to capture long-range dependencies makes it promising for application to financial time series data. This paper chooses to employ the LSTM model with 3 timesteps, which means we incorporate the SPY call transactions at time $t-2$, $t-1$, and $t$.\footnote{$t-2$, $t-1$, and $t$ represent the three closest days in which eligible options are traded, and they don't necessarily occur on three consecutive days.} 
Before training on the time sequences, we need to reselect and extract data from the 15 datasets. We take the features of SPY call records from the original samples with the closest three trading days that share the same strike price. In total, 18 features are chosen as the features for the new time series group, with the target output being the SPY call price at time $t$. Through this filtering method, we can observe the influence of historical option features on the latest option price for options with the same strike, as well as reveal the correlations between various features in the time series.
In addition, we intend to further investigate the impact of call prices of historical moments on the call price at time $t$. Therefore, we attempt to include historical call prices as features in our LSTM model, and the final set of 21 features consists of: 
\textit{call price1}, \textit{spot price1}, \textit{strike1}, \textit{moneyness1}, \textit{days to maturity1}, \textit{volatility1}, \textit{interest rate1}; 
\textit{call price2}, \textit{spot price2}, \textit{strike2}, \textit{moneyness2}, \textit{days to maturity2}, \textit{volatility2}, \textit{interest rate2}; 
\textit{call price3}, \textit{spot price3}, \textit{strike3}, \textit{moneyness3}, \textit{days to maturity3}, \textit{volatility3}, \textit{interest rate3}.\footnote{The index $1$ denotes the latest time and indices $2$ and $3$ present the historic times. To ensure the same length of feature group for the three timesteps, \textit{call price1} is set to 0 for all instances.}
After the filtering process, the count of records in the chosen time series groups reduces from the initial 2,794,966 to 188,049 which is still ample for training the neural networks. However, we tailor the number of training and validation epochs for the time series data instead of 2000 times before. In specific, we divide 200,000,000 by the number of training data in each of the 15 datasets to determine the number of epochs respectively. Additionally, we stop training and save the current "best" trained model if the loss of validation set remains unchanged for a consecutive 2000 epochs.\footnote{It is known as the early stopping technique in machine learning training.} 
We design it this way because with a reduced sample size, increasing the number of epochs allows the model to thoroughly explore the inherent relationships within the data. Yet, if the loss does not decrease for an extended period, it indicates convergence has been achieved.

In order to grasp time-series insights and their varying influence on input features and time lags, we try to combine the LSTM architecture with the self-attention mechanism. The self-attention layer is added between the six LSTM layers and the fully connected layer. Due to the limited count of our time series data, we consider replacing the LSTM architecture with the GRU architecture. The GRU structure has fewer parameters, resulting in faster training speed, and it tends to perform better for induction with smaller datasets compared to LSTM. The self-attention GRU model is built with six GRU layers, one self-attention layer, and one fully connected layer before the final output. Other hyperparameters of the self-attention GRU model keep the same as the LSTM neural network. 

This paper utilizes three metrics to assess the performance of various models. Their specific calculation methods are as follows:
\begin{itemize}
    \item \textbf{Root Mean Squared Error (RMSE):}
\begin{equation}
RMSE = \sqrt{\frac{1}{n} \sum_{i=1}^{n} (y_i - \hat{y}_i)^2}
\end{equation}
    \item \textbf{Mean Absolute Percentage Error (MAPE):}
\begin{equation}
MAPE = \frac{1}{n} \sum_{i=1}^{n} \left| \frac{y_i - \hat{y}_i}{y_i} \right| \times 100
\end{equation}
    \item \textbf{Coefficient of Determination (R-squared):}
\begin{equation}
R^2 = 1 - \frac{\sum_{i=1}^{n} (y_i - \hat{y}_i)^2}{\sum_{i=1}^{n} (y_i - \bar{y})^2}
\end{equation}
\end{itemize}
where $n$ is the number of samples, $y_i$ represents the actual values, $\hat{y}_i$ is the predicted values, and $\bar{y}$ is the mean of the actual values.

\begin{table*}[h]
\centering
\caption{RMSE of models' predictions on test data}
\begin{tabular}{p{1.5cm}p{1.5cm}p{1.5cm}p{1.5cm}p{1.5cm}p{1.55cm}p{1.55cm}p{1.7cm}p{1.7cm}}
\toprule
Moneyness          & Maturity          & BT      & MLP    & LSTM\_6F   & LSTM\_18F    & LSTM\_21F    & SA\_LSTM\_21F & SA\_GRU\_21F \\ \midrule
\multirow{5}{*}{$\leq$0.97}    & 1-9   & 0.6677  & 0.0487 & 0.0233  & 0.0075*  & 0.0098**   & 0.0156    & 0.0117   \\
                   & 10-30             & 2.3835  & 0.0874 & 0.0239  & 0.0293   & 0.0217** & 0.0210*   & 0.0218   \\
                   & 31-90             & 5.5773  & 0.1706 & 0.0540  & 0.0733   & 0.0559   & 0.0481*   & 0.0508** \\
                   & 91-180            & 11.2895 & 0.2522 & 0.1186  & 0.1063   & 0.0971*  & 0.1070    & 0.1008**   \\
                   & \textgreater{}180 & 23.7380 & 0.4796 & 0.2714* & 0.2833   & 0.2915   & 0.3174    & 0.2829** \\
\midrule
\multirow{5}{*}{0.97$\sim$1.03}& 1-9   & 1.4329  & 0.3361 & 0.2547  & 0.0827** & 0.0913   & 0.0819*   & 0.0828   \\
                   & 10-30             & 3.4530  & 0.3863 & 0.1514  & 0.0872   & 0.0867   & 0.0755**  & 0.0723*  \\
                   & 31-90             & 6.9494  & 0.5198 & 0.1882  & 0.1301   & 0.1233   & 0.1103**  & 0.1032*  \\
                   & 91-180            & 12.7069 & 0.5154 & 0.3272  & 0.3150   & 0.2070   & 0.1829*    & 0.1929**   \\
                   & \textgreater{}180 & 20.7043 & 0.6336 & 0.3518  & 0.3987   & 0.3552   & 0.3280**    & 0.2802*  \\
\midrule
\multirow{5}{*}{\textgreater{}1.03}&1-9& 0.3466  & 0.2631 & 0.2486  & 0.1769   & 0.1754**   & 0.1844    & 0.1345*  \\
                   & 10-30             & 0.7867  & 0.3440 & 0.2930  & 0.2446   & 0.2196**   & 0.2321    & 0.1428*  \\
                   & 31-90             & 2.1018  & 0.4170 & 0.3395  & 0.2420   & 0.2168   & 0.1935**    & 0.1550*  \\
                   & 91-180            & 4.7588  & 0.4796 & 0.3498  & 0.5335   & 0.3040   & 0.2555**  & 0.2401*  \\
                   & \textgreater{}180 & 11.2180 & 0.6895 & 0.3194  & 0.3902   & 0.3445   & 0.3250**  & 0.2598*  \\ 
\bottomrule
\end{tabular}
\label{table1}
\end{table*}

\begin{table*}[h]
\centering
\caption{MAPE of models' predictions on test data}
\begin{tabular}{p{1.5cm}p{1.5cm}p{1.5cm}p{1.5cm}p{1.5cm}p{1.55cm}p{1.55cm}p{1.7cm}p{1.7cm}}
\toprule
Moneyness          & Maturity          & BT      & MLP    & LSTM\_6F   & LSTM\_18F    & LSTM\_21F    & SA\_LSTM\_21F & SA\_GRU\_21F \\ \midrule
\multirow{5}{*}{$\leq$0.97} & 1-9 & 447.4012 & 61.2574 & 37.0200 & 19.2545** & 19.9389 & 23.9054  & 18.5316* \\
 & 10-30 & 1351.6848 & 39.0816 & 18.4638 & 10.5836 & 8.2162*  & 10.3189** & 11.5799 \\
 & 31-90 & 1826.4752 & 62.6888 & 16.1679 & 8.1250** & 11.4298 & 7.9859* & 9.6918 \\
 & 91-180 & 2102.9330 & 38.8356 & 11.9290 & 8.6297 & 5.9395**& 8.5348 & 5.5468* \\
 & >180 & 1252.0830 & 14.7654 & 6.6863 & 6.4090** & 6.8350  & 9.0688  & 6.3777* \\
 \midrule
\multirow{5}{*}{0.97$\sim$1.03} & 1-9 & 157.5662 & 25.9497 & 17.6635 & 5.5764* & 6.5916  & 6.5399  & 6.1179** \\
 & 10-30 & 104.5943 & 7.4550 & 2.7668 & 1.2787 & 0.9166**  & 1.0988 & 0.8792* \\
 & 31-90 & 84.0531 & 4.2464 & 1.6238 & 0.9380 & 0.7924  & 0.7668** & 0.6153* \\
 & 91-180 & 71.6790 & 2.2567 & 1.4141 & 1.2183 & 0.7352*  & 0.7409** & 0.7787 \\
 & >180 & 68.8223 & 1.6229 & 0.8437 & 0.9178 & 0.8894  & 0.8391**  & 0.6942** \\
 \midrule
\multirow{5}{*}{>1.03} & 1-9 & 0.6635 & 0.4892 & 0.4578 & 0.3244** & 0.3381 & 0.3489  & 0.2383* \\
 & 10-30 & 1.7553 & 0.6752 & 0.5571 & 0.5115 & 0.4246** & 0.4404  & 0.2588* \\
 & 31-90 & 4.0971 & 0.6639 & 0.5089 & 0.4399 & 0.3474 & 0.3279**  & 0.2143* \\
 & 91-180 & 7.5876 & 0.6155 & 0.4180 & 0.4487 & 0.2378** & 0.2706 & 0.1789* \\
 & >180 & 12.8379 & 0.6619 & 0.2658** & 0.4773 & 0.3844 & 0.4222  & 0.2409* \\ 
\bottomrule
\end{tabular}
\label{table2}
\end{table*}

\begin{table*}[h]
\centering
\caption{R-squared of models' predictions on test data}
\begin{tabular}{p{1.5cm}p{1.5cm}p{1.5cm}p{1.5cm}p{1.5cm}p{1.55cm}p{1.55cm}p{1.7cm}p{1.7cm}}
\toprule
Moneyness          & Maturity          & BT      & MLP    & LSTM\_6F   & LSTM\_18F    & LSTM\_21F    & SA\_LSTM\_21F & SA\_GRU\_21F \\ \midrule
\multirow{5}{*}{$\leq$0.97} & 1-9 & -1.854700 & 0.982250 & 0.995945 & 0.999229* & 0.998701** & 0.996679 & 0.998140 \\
 & 10-30                          & -3.862500 & 0.993671 & 0.999528 & 0.999504 & 0.999726**  & 0.999744* & 0.999724 \\
 & 31-90                          & -5.527843 & 0.993618 & 0.999361 & 0.999144 & 0.999502 &  0.999632* & 0.999588** \\
 & 91-180                         & -6.265260 & 0.996315 & 0.999185 & 0.999626 & 0.999688* & 0.999621 &  0.999664** \\
 & >180                           & -5.975940 & 0.997192 & 0.999101 & 0.999152** & 0.999102  & 0.998935 & 0.999154* \\
 \midrule
 \multirow{5}{*}{0.97$\sim$1.03} & 1-9 & 0.861986 & 0.992414 & 0.995643 & 0.999483** & 0.999369  & 0.999492* & 0.999481 \\
 & 10-30                               & 0.402577 & 0.992440 & 0.998839 & 0.999628 & 0.999632 &  0.999721** & 0.999745* \\
 & 31-90                               & -0.826653 & 0.989849 & 0.998670 & 0.999385 & 0.999447 & 0.999558** & 0.999613* \\
 & 91-180                              & -4.029520 & 0.991721 & 0.996663 & 0.997707 & 0.999009  & 0.999227*  & 0.999139** \\
 & >180                                & -3.381820 & 0.995936 & 0.998747 & 0.998202 & 0.998573 & 0.998783**  & 0.999112* \\
 \midrule
 \multirow{5}{*}{>1.03} & 1-9 & 0.999958 & 0.999976 & 0.999978 & 0.999989** & 0.999989** & 0.999988   & 0.999994* \\
 & 10-30                      & 0.999782 & 0.999958 & 0.999970 & 0.999977 & 0.999982** & 0.999980  & 0.999992* \\
 & 31-90                      & 0.998805 & 0.999953 & 0.999969 & 0.999980 & 0.999984 & 0.999987** & 0.999992* \\
 & 91-180                     & 0.994474 & 0.999944 & 0.999970 & 0.999937 & 0.999980 & 0.999986*  & 0.999987* \\
 & >180                       & 0.974110 & 0.999903 & 0.999979* & 0.999943 & 0.999955 & 0.999960 & 0.999975** \\
\bottomrule
\end{tabular}
\label{table3}
\end{table*}

\subsubsection{Model Comparison}
The performance of different models evaluated using RMSE, MAPE, and R-squared is presented in Tables \ref{table1}, \ref{table2}, and \ref{table3}, respectively. The abbreviations "BT", "MLP", "LSTM\_6F", "LSTM\_18F", "LSTM\_21F", "SA\_LSTM\_21F", and "SA\_GRU\_21F" correspond to the binomial tree model, MLP model, LSTM model with 6 features, LSTM model with 18 features, LSTM model with 21 features, self-attention LSTM model with 21 features, and self-attention GRU model with 21 features. The sign "$*$" denotes the best performance among all models sharing the same moneyness and maturity, and "$**$" presents the second-best performance.

Firstly, as listed in Tables \ref{table1} to \ref{table3}, it is evident that the predictive performance of the deep learning models is significantly better than that of the traditional binomial tree model. Even the most fundamental deep learning model, the MLP, exhibits a substantial improvement compared to the binomial tree model. In terms of the RMSE metric, the maximum enhancement goes from 23.7380 to 0.4796; for the MAPE metric, the largest improvement is from 2102.9330 to 38.8356; and for the R-squared metric, the highest improvement is from -6.265260 to 0.996315. In other words, the most basic MLP neural network achieves an improvement of over 50 times in accuracy compared to the traditional pricing model for American options. Moreover, the computational time of neural network models is much lower than that of the binomial tree. This is because the binomial tree pricing method is path-dependent, requiring considerable time for underlying path fitting. In contrast, neural network models can be further accelerated using GPU to enhance computational efficiency. 

When considering moneyness classification, the binomial tree model demonstrates the most accurate predictions for ITM options and the least accurate predictions for OTM options. Afterwards, transitioning from the binomial tree model to the MLP model and then to the LSTM model without historical data (i.e., the LSTM model
with 6 features), there is a notable improvement in the predictive accuracy of options across the three categories of moneyness. The LSTM model with 18 features displays an overall improvement in predictive capability in contrast to the LSTM model with 6 features, suggesting that historical features of options contribute to predicting current call prices. Intuitively, historical call prices should aid in predicting current call prices, and we indeed prove this guess through our experimentation: the LSTM model with 21 features outperforms the LSTM model with 18 features, especially for the ATM and ITM options. 

\begin{figure*}
\centering
\subfigure[RMSE of OTM options sorted by maturity]{
\includegraphics[width=8.53cm]{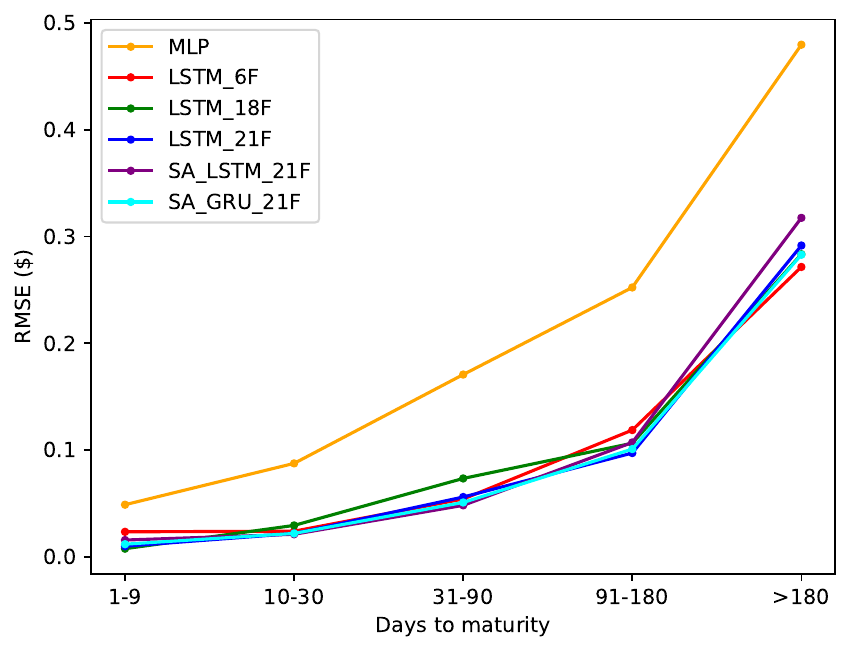}
}
\quad
\subfigure[MAPE of OTM options sorted by maturity]{
\includegraphics[width=8.53cm]{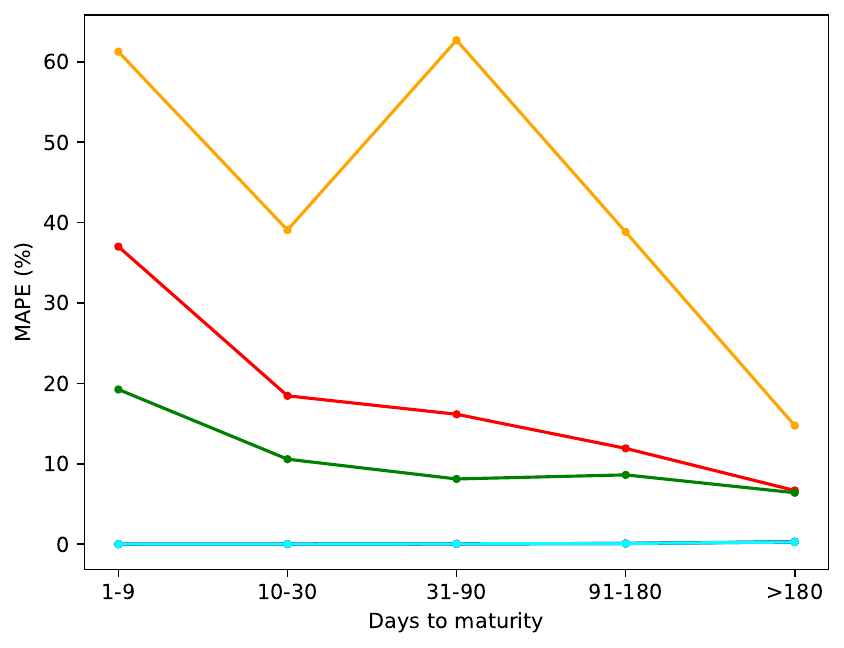}
}
\quad
\subfigure[RMSE of ATM options sorted by maturity]{
\includegraphics[width=8.53cm]{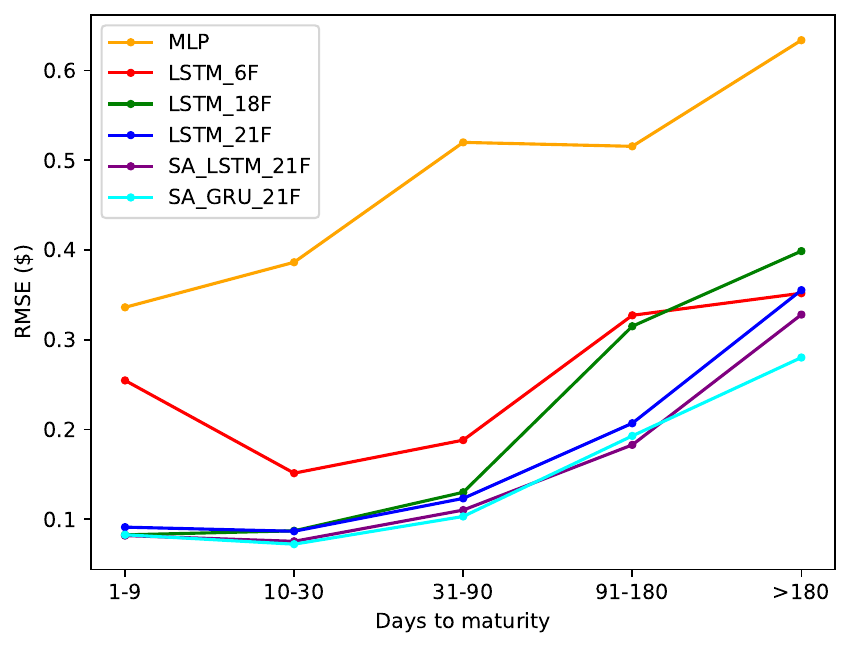}
}
\quad
\subfigure[MAPE of ATM options sorted by maturity]{
\includegraphics[width=8.53cm]{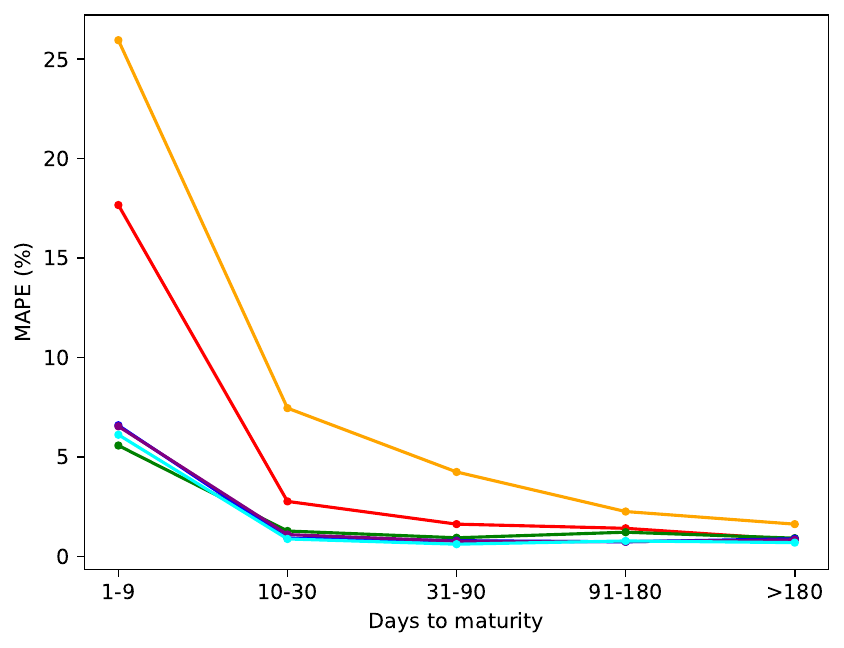}
}
\quad
\subfigure[RMSE of ITM options sorted by maturity]{
\includegraphics[width=8.53cm]{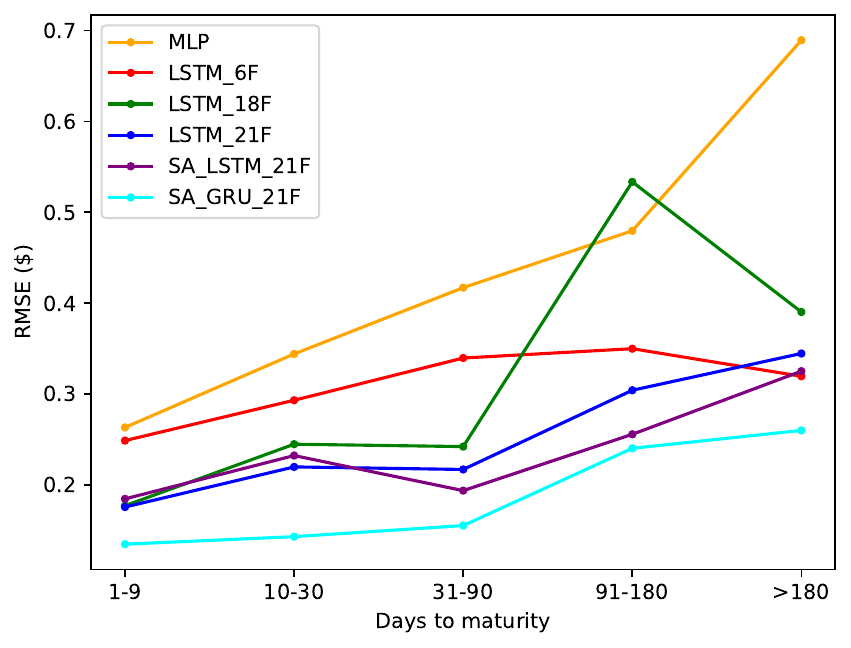}
}
\quad
\subfigure[MAPE of ITM options sorted by maturity]{
\includegraphics[width=8.53cm]{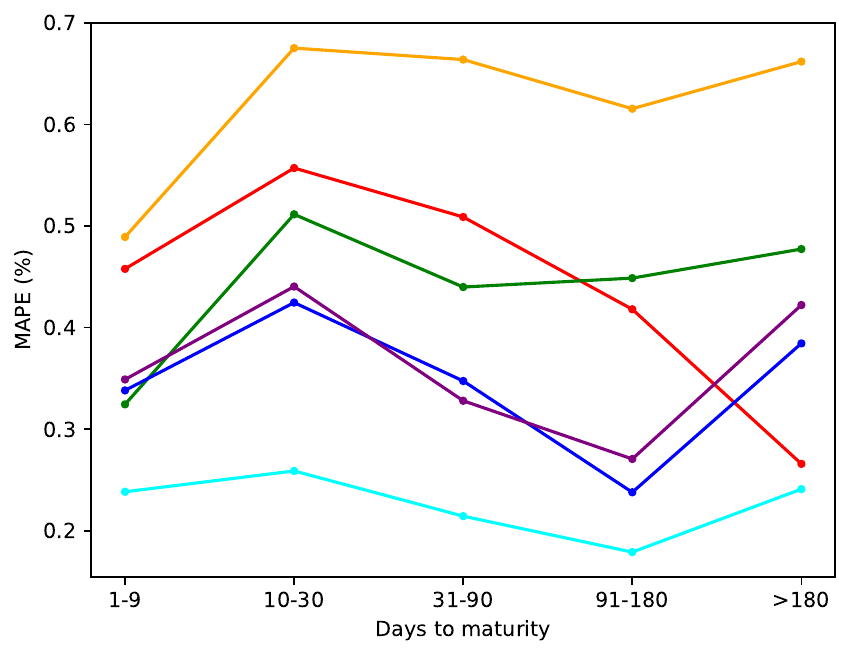}
}
\caption{Comparison of models' performance}
\label{Comparison of models' performance}
\end{figure*}

Finally, we conduct experiments on the combinations of self-attention with LSTM and GRU correspondingly, both of them based on the 21 features. 
It turns out that the accuracies of ATM and ITM options continue to enhance with self-attention and the improvement is relatively less pronounced for the OTM type. 
The whole experimental process for our models is a step-by-step refinement of accuracy. 
Ultimately, the results indicate that the self-attention GRU model with 21 features is the best-performing pricing approach for the SPY call options. 
This can be visually observed from the Figure \ref{Comparison of models' performance}.\footnote{Due to the significant disparity in accuracy between the binomial tree and deep learning models, we opt not to include the performance of the binomial tree in the figures.}

\subsection{Interpretable analysis}
In order to make the deep learning model understandable and reliable, we employ the SHapley Additive exPlanations (SHAP) method to analyze our best-performing model, the self-attention GRU model with 21 features. We apply the algorithms proposed by Lundberg and Lee\footnote{https://github.com/shap/shap} and the SHAP summary plots of 15 datasets are depicted in Figure \ref{SHAP plots for options sorted by moneyness and maturity}.

The SHAP value provides the average feature contribution to the self-attention GRU model with 21 features classified by moneyness and days to maturity. The SHAP summary plot presents feature importance in a descending order from top to bottom. This ranking is determined by calculating the mean of the absolute SHAP values for each feature. Overall, the three features that have the greatest impact on the final predicted call price are: \textit{spot price1}, \textit{strike1}, and \textit{moneyness1}. Among them, \textit{spot price1} and \textit{moneyness1} have a positive impact, while \textit{strike1} has a negative impact on the prediction. This phenomenon makes sense since American call options with higher spot prices and lower strike prices (i.e., higher moneyness) tend to have larger intrinsic values and are more likely to be early exercised. As a result, their current option prices are likely to be higher. Other more detailed analysis results associated with moneyness and days to maturity are as follows:
\begin{itemize}
    \item The feature importance ranking of ATM options is different from those for OTM and ITM types. Particularly, \textit{moneyness1} ranks first in terms of its average contribution, while \textit{spot price1} and \textit{strike1} do not exhibit significant impacts on the predicted call price. The special difference illustrates that the moneyness of ATM options can have a greater impact on option prices compared to the individual spot price and strike price. ATM options have little intrinsic value, primarily consisting of time value, which is closely related to the moneyness factor. Besides, moneyness as a dynamic concept is more susceptible to change along with shifts in market conditions and other macroeconomic factors.
    \item Both \textit{call price2} and \textit{call price3} have significant positive impacts in OTM, ATM, and ITM options, with the impact of \textit{call price2} being particularly pronounced in the case of ATM options (i.e., four out of five sampled ATM options rank it as the second-highest factor). The result shows that past call prices have a favorable effect on forecasting current call prices. This effect is particularly noticeable in ATM options, which are more prone to being influenced by the nearest historical call prices.
    \item ITM options receive substantial positive influence from \textit{spot price2} and experience significant negative impact from \textit{spot price3}. From this, it can be inferred that \textit{spot price1} is positively correlated with \textit{spot price2} but negatively correlated with \textit{spot price3}. We speculate that the reason is the relatively longer time interval between \textit{spot price1} and \textit{spot price3}, which increases the likelihood of underlying price fluctuations occurring. \textit{Spot price3} similarly exerts a detrimental influence on OTM options, while the impact of \textit{spot price2} on such options is relatively modest. However, after the days to maturity exceed 30 days, \textit{spot price2} exhibits a positive impact on the predicted values. 
    \item Both \textit{strike2} and \textit{strike3} have substantial negative effects on ITM options, consistent with the pattern observed with \textit{strike1}. As for the OTM options, \textit{strike2} negatively impacts the prediction of call prices, whereas \textit{strike3} has a positive effect except when the maturity is less than 9 days. The positive influence of \textit{strike3} on the forecasted values of OTM options contradicts the intuition, as all three strikes within the same time series group are identical. Nonetheless, considering the negative compact of \textit{spot price3} mentioned earlier, this makes the positive impact of \textit{strike3} less surprising. 
    \item \textit{Moneyness2} shows a positive effect on both OTM and ITM options (similar to moneyness1), while \textit{moneyness3} presents a negative impact. By contrast, these two features do not exhibit clear directional influences on the ATM options. We can conclude that ATM options are more distinctly positively influenced by \textit{moneyness1}, whereas other features such as \textit{moneyness2}, \textit{moneyness3}, \textit{spot price1}, \textit{spot price2}, \textit{spot price3}, \textit{strike1}, \textit{strike2}, and \textit{strike3} do not display clear patterns in the final predictions.
    \item The feature \textit{volatility1} exerts a positive influence on all three types of options, but with varying magnitudes of impact: it has a greater impact on OTM and ATM options, while its influence is weaker on ITM options (gradually turning positive for maturities beyond 90 days). As the SPY price experiences greater volatility, the likelihood of an increase in the spot price rises. This is a desired outcome for OTM and ATM call options. Therefore, underlying volatility has a positive effect on the predicted call prices of these options. However, for ITM options, where the spot price is already higher than the strike price, they do not favor high underlying volatility in the short term. Nevertheless, as the expiration date extends (beyond 90 days), the potential for further spot price increases rises. Consequently, \textit{volatility1} gradually presents a positive influence on the predicted prices of ITM options over time.
    \item \textit{Volatility2} and \textit{volatility3} have relatively minor impacts on OTM, ATM, and ITM options, and overall exhibit negative contributions to the final predictions. ATM options are most affected by these two features among the three categories. The negative relationship between the volatility of the historical underlying asset and the model output indicates that the current call option prices do not favor past fluctuations in the underlying prices.
    \item The feature \textit{days to maturity1} contributes positively to the final predictions of ATM and OTM options, especially for the ATM type. ATM options have minimal intrinsic value and are primarily composed of time value, making the remaining time before expiration crucial for them. In contrast, ITM options derive a main portion of their value from intrinsic value, which is why their sensitivity to the \textit{days to maturity1} is not as pronounced, although it remains positive. \textit{Days to maturity2} and \textit{days to maturity3} both present unclear trends and have limited effects on all three types of options.
    \item The influence of risk-free interest rates on option predictions is minor,  yet there is still a discernible pattern. For OTM options with days to maturity exceeding 9 days, \textit{interest rate1} has a positive impact, while for ATM and ITM options with days to maturity exceeding 30 days, it also has a positive impact on the predictive call price. We can deduce that as the maturity of options extends, a higher matched risk-free interest rate tends to have a positive influence on the model's results. Conversely, when the remaining days before expiry are shorter, the current interest rate has a negative impact on the prediction. Generally, when the risk-free interest rate increases, it results in an increase in the expected return on the underlying asset of the options. Additionally, the rise in the risk-free interest rate, which serves as the discount rate, leads to a decrease in the present value of future earnings for option holders. For call options, the two effects work in opposite directions: the first effect causes the option price to rise, while the second effect causes it to fall.
    With shorter days to maturity, the second effect dominates, causing an increase in the risk-free interest rate to lead to a decrease in option prices. On the other hand, with longer days to maturity, the first effect prevails, causing an increase in the risk-free interest rate to result in higher option prices. However,  in the broader context, the effect of interest rate on the forecasted call prices is quite modest when contrasted with other features.
\end{itemize}

\begin{figure*}
\centering
\subfigure[Moneyness $\leq$ 0.97 \& Maturity $\leq$ 9 days]{
\includegraphics[width=8.53cm]{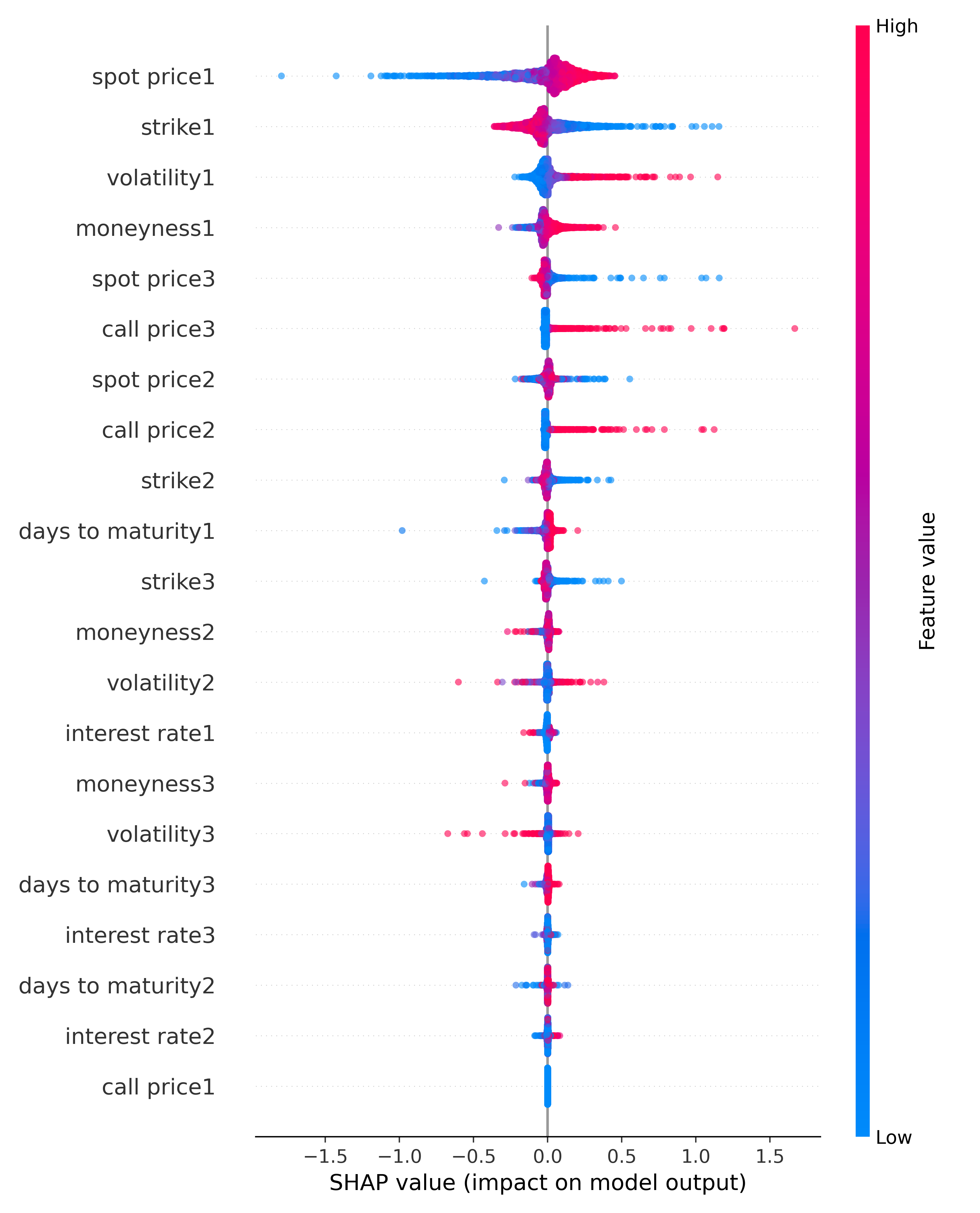}
}
\quad
\subfigure[Moneyness $\leq$ 0.97 \& 9 days $<$ Maturity $\leq$ 30 days]{
\includegraphics[width=8.53cm]{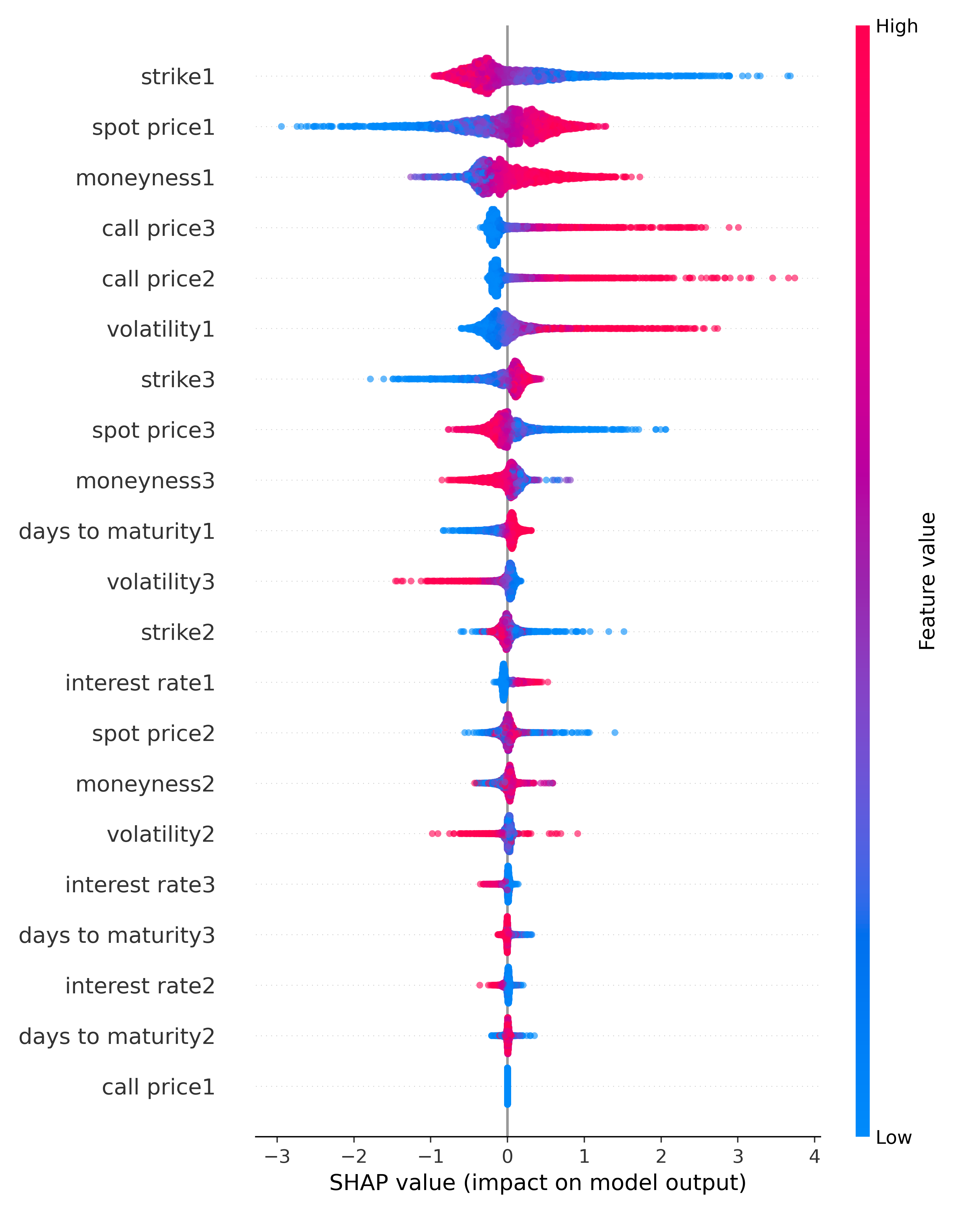}
}
\quad
\subfigure[Moneyness $\leq$ 0.97 \& 30 days $<$ Maturity $\leq$ 90 days]{
\includegraphics[width=8.53cm]{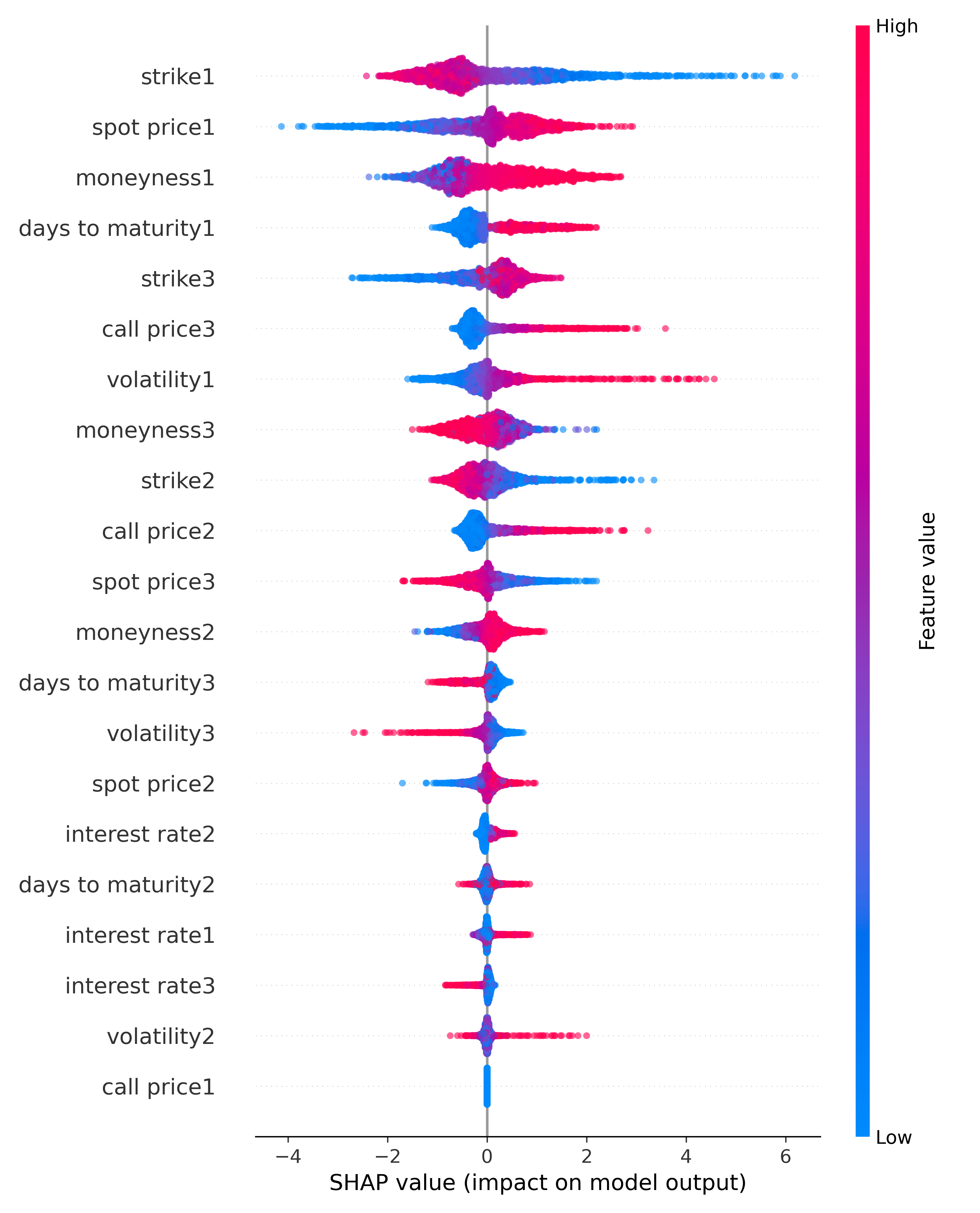}
}
\quad
\subfigure[Moneyness $\leq$ 0.97 \& 90 days $<$ Maturity $\leq$ 180 days]{
\includegraphics[width=8.53cm]{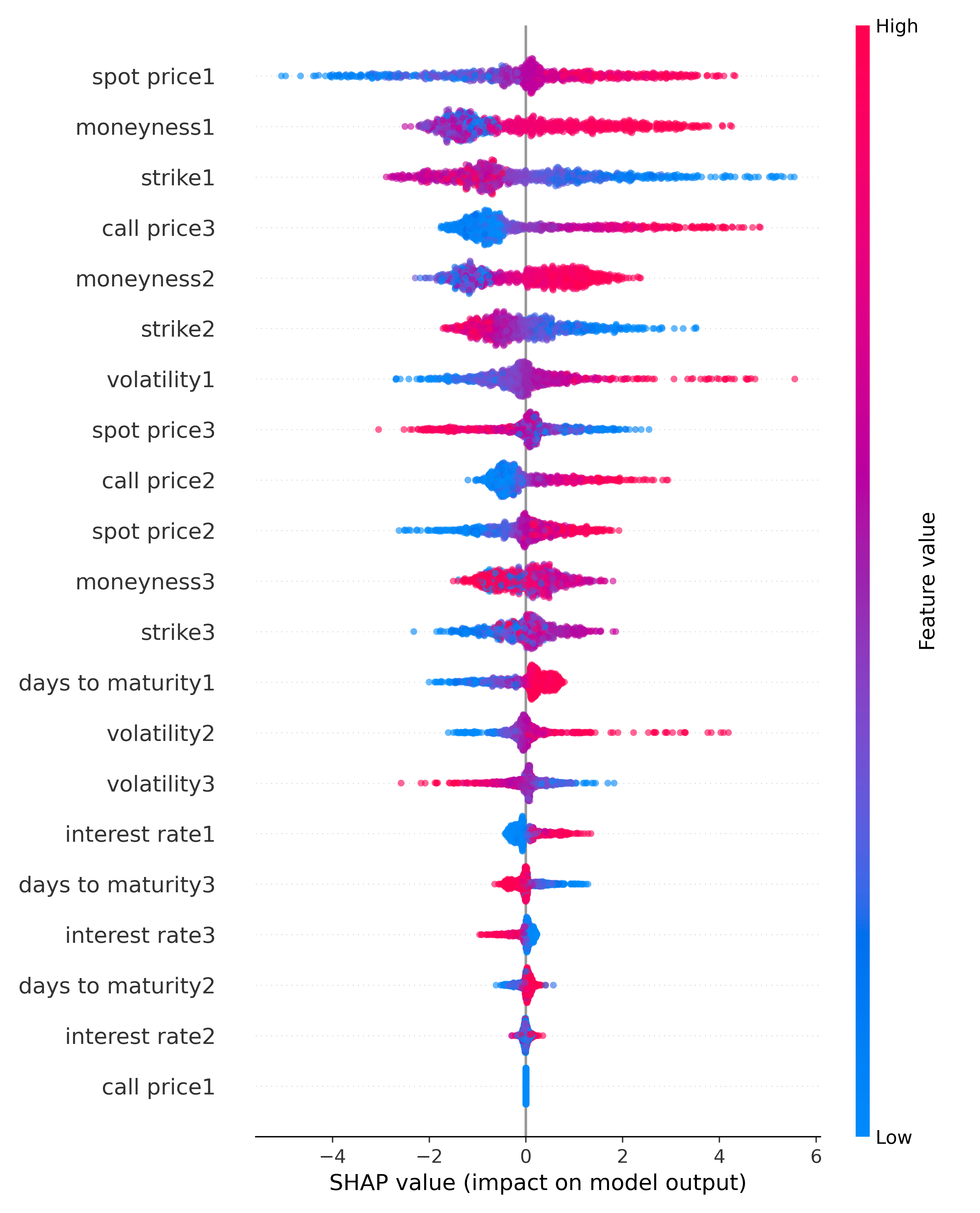}
}
\caption{SHAP plots for options sorted by moneyness and maturity}
\label{SHAP plots for options sorted by moneyness and maturity}
\end{figure*}

\begin{figure*}\ContinuedFloat
\centering
\addtocounter{subfigure}{+4}
\subfigure[Moneyness $\leq$ 0.97 \& Maturity $>$ 180 days]{
\includegraphics[width=8.53cm]{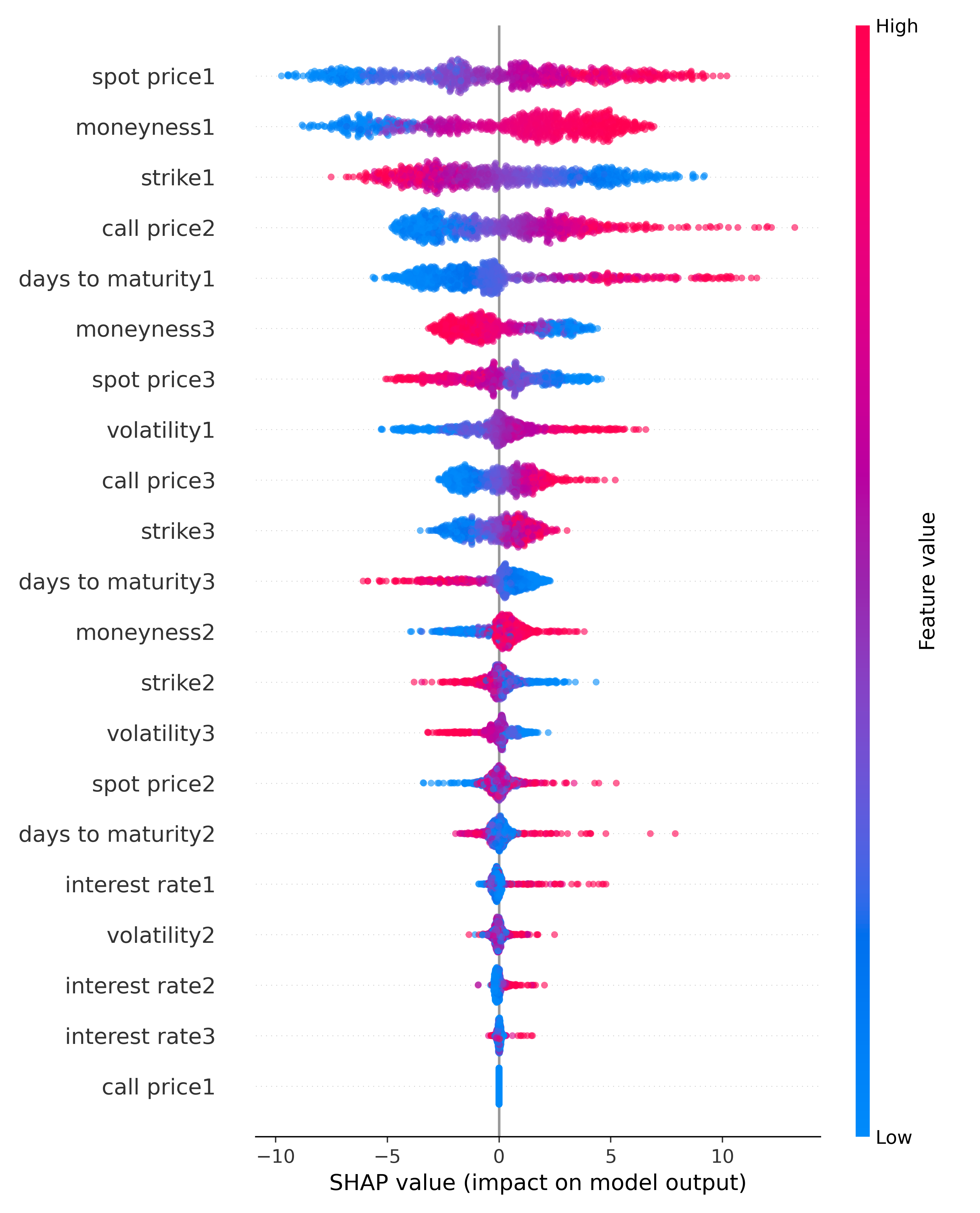}
}
\quad
\subfigure[0.97 $<$ Moneyness $\leq$ 1.03 \& Maturity $\leq$ 9 days]{
\includegraphics[width=8cm]{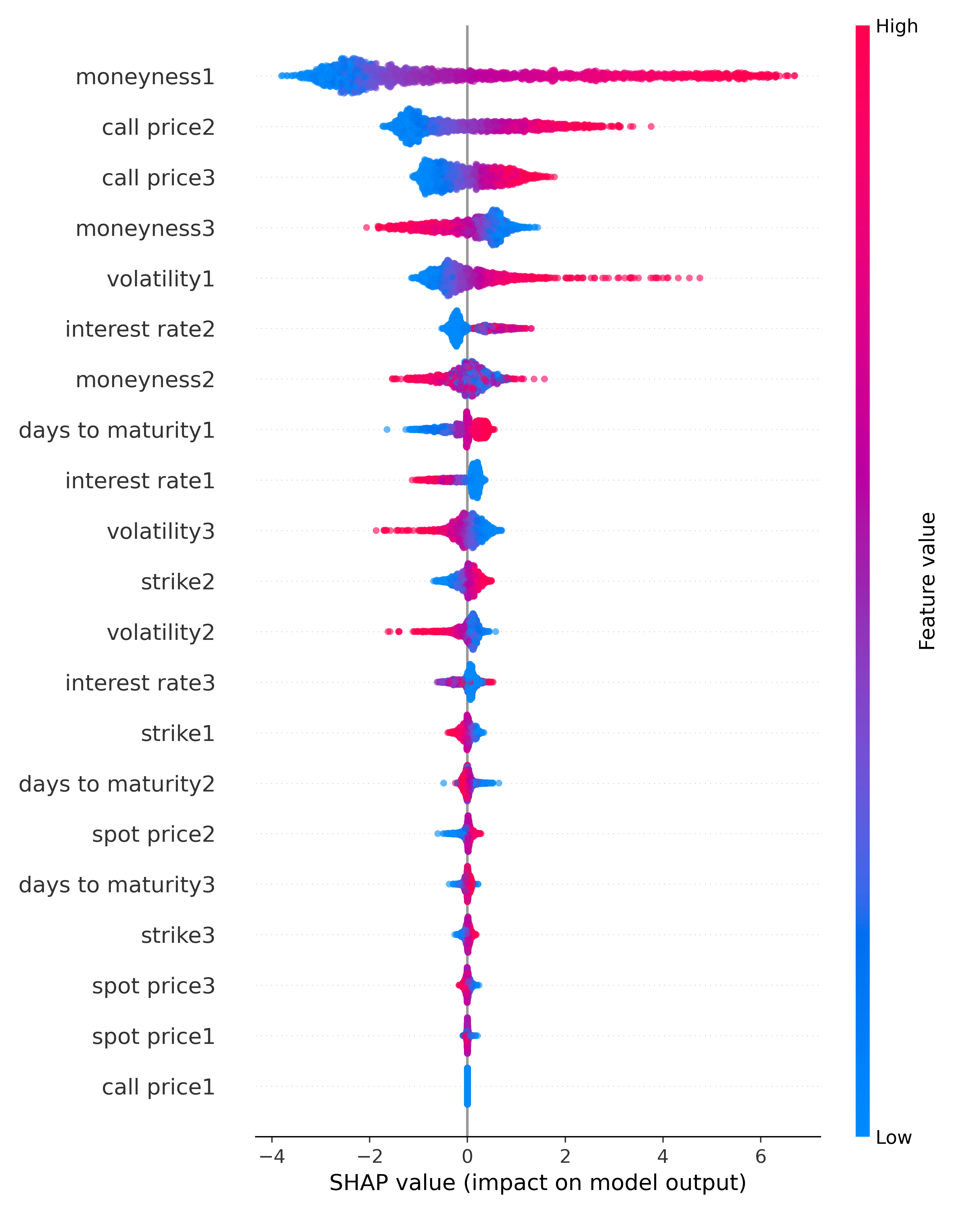}
}
\quad
\subfigure[0.97 $<$ Moneyness $\leq$ 1.03 \& 9 days $<$ Maturity $\leq$ 30 days]{
\includegraphics[width=8cm]{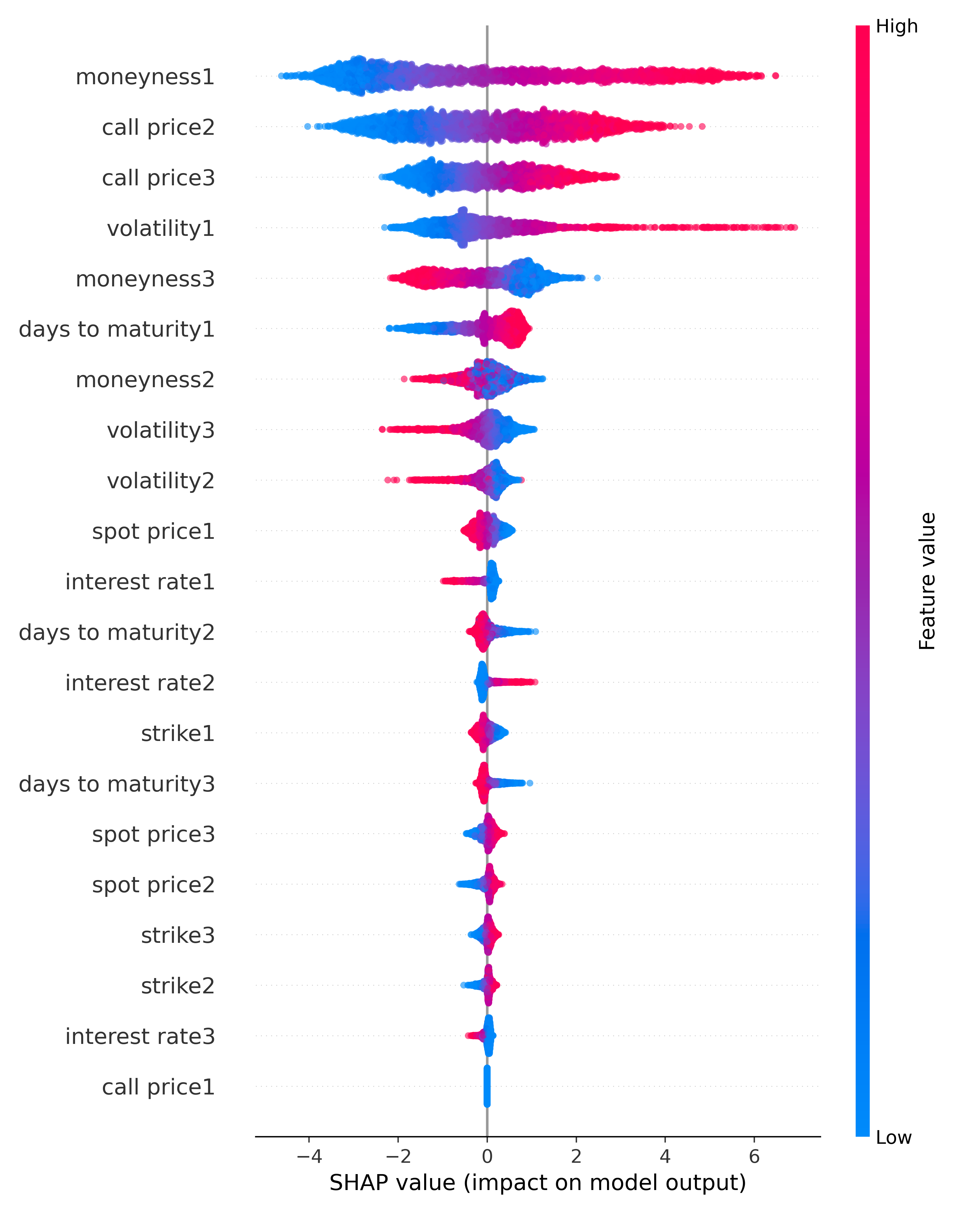}
}
\quad
\subfigure[0.97 $<$ Moneyness $\leq$ 1.03 \& 30 days $<$ Maturity $\leq$ 90 days]{
\includegraphics[width=8.53cm]{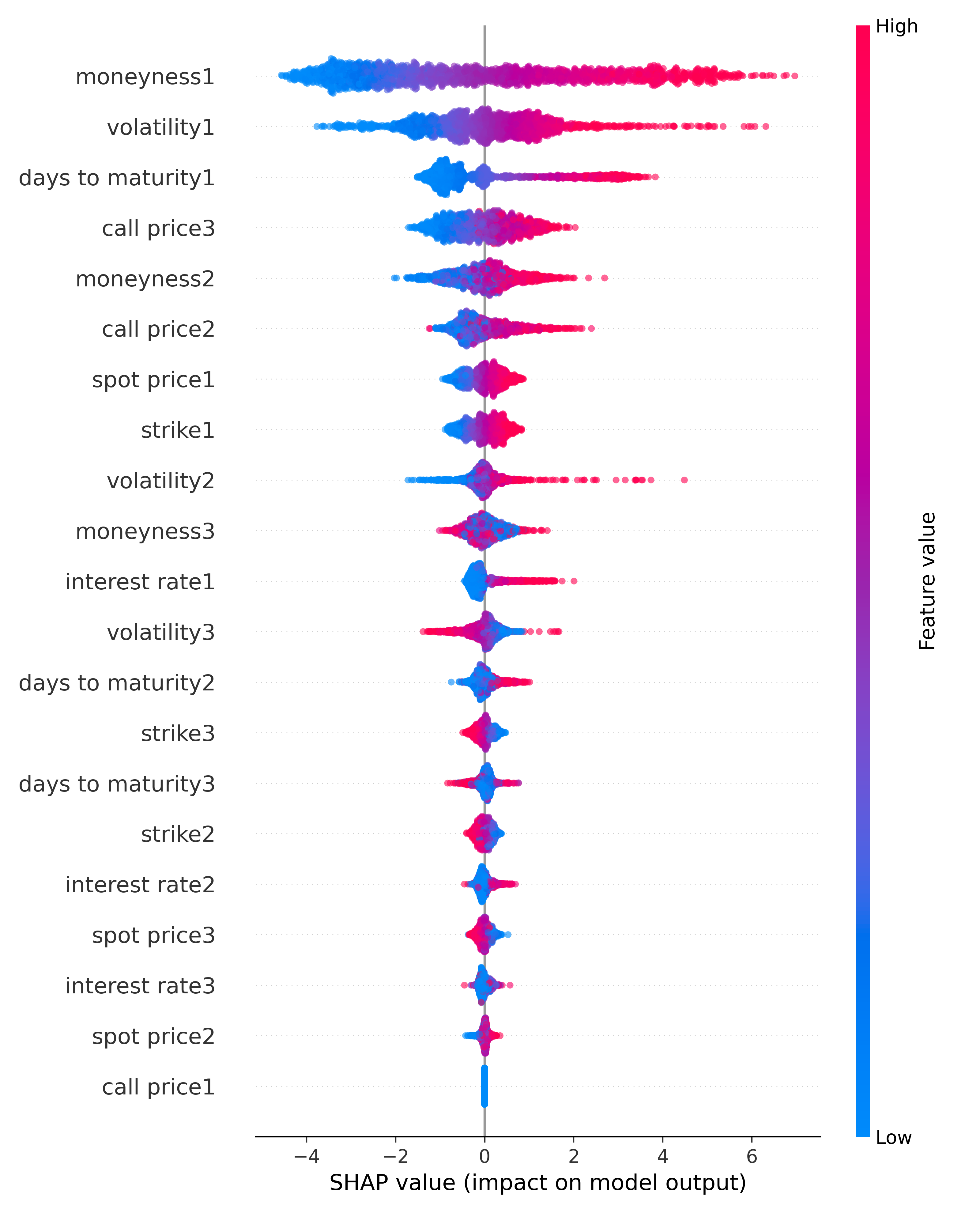}
}
\caption{SHAP plots for options sorted by moneyness and maturity}
\label{SHAP plots for options sorted by moneyness and maturity}
\end{figure*}

\begin{figure*}\ContinuedFloat
\centering
\subfigure[0.97 $<$ Moneyness $\leq$ 1.03 \& 90 days $<$ Maturity $\leq$ 180 days]{
\includegraphics[width=8.53cm]{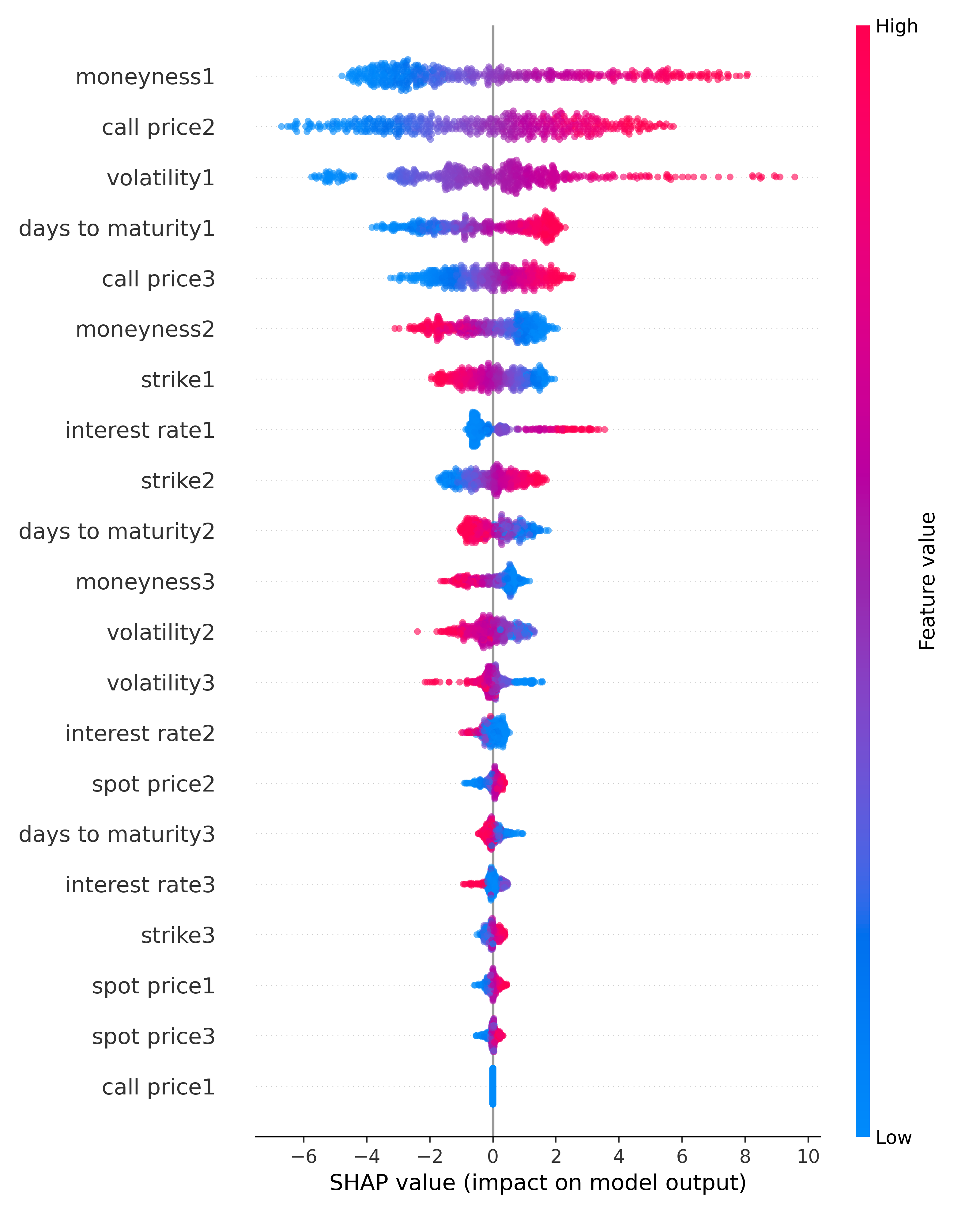}
}
\quad
\subfigure[0.97 $<$ Moneyness $\leq$ 1.03 \& Maturity $>$ 180 days]{
\includegraphics[width=8.53cm]{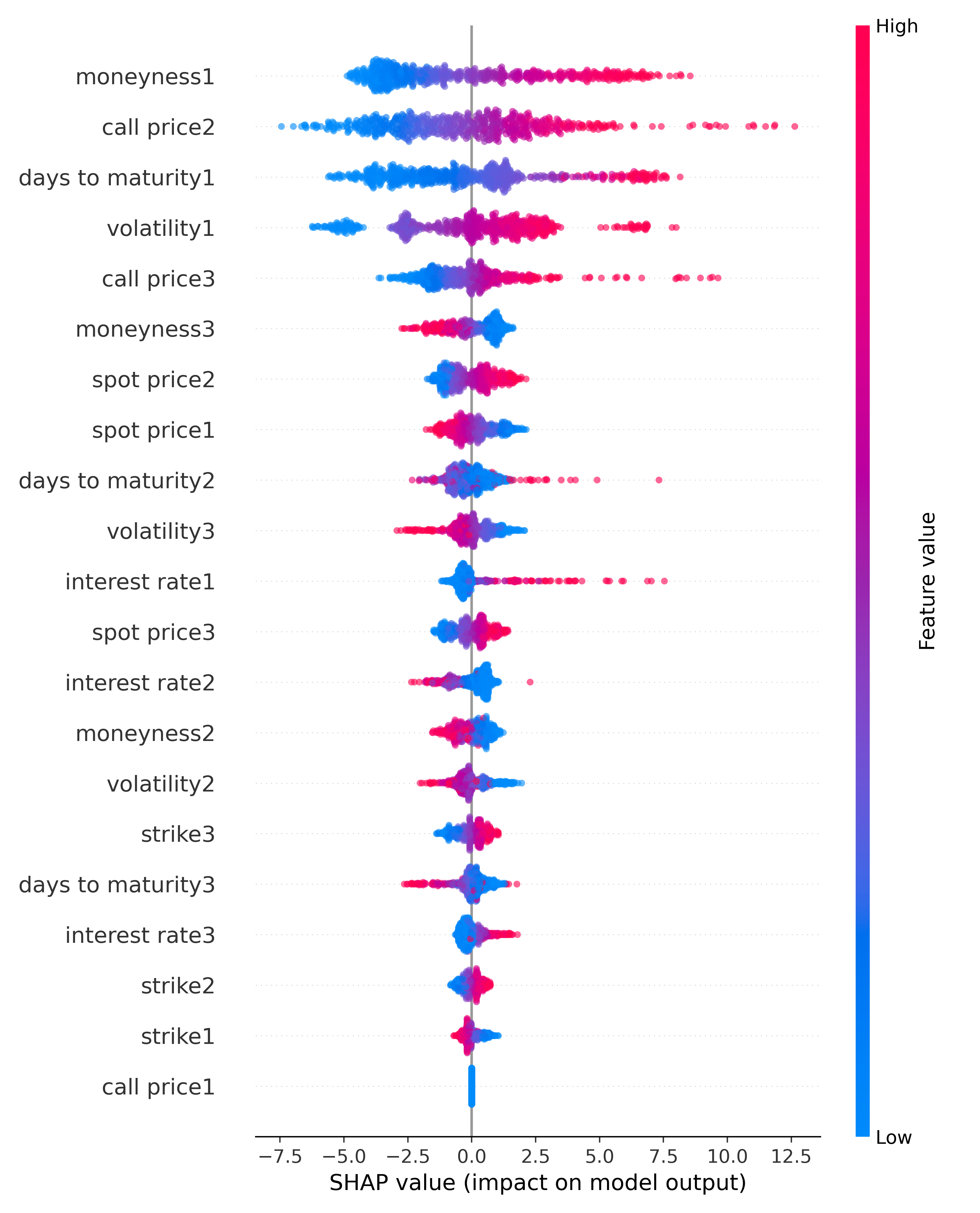}
}
\quad
\subfigure[Moneyness $>$ 1.03 \& Maturity $\leq$ 9 days]{
\includegraphics[width=8cm]{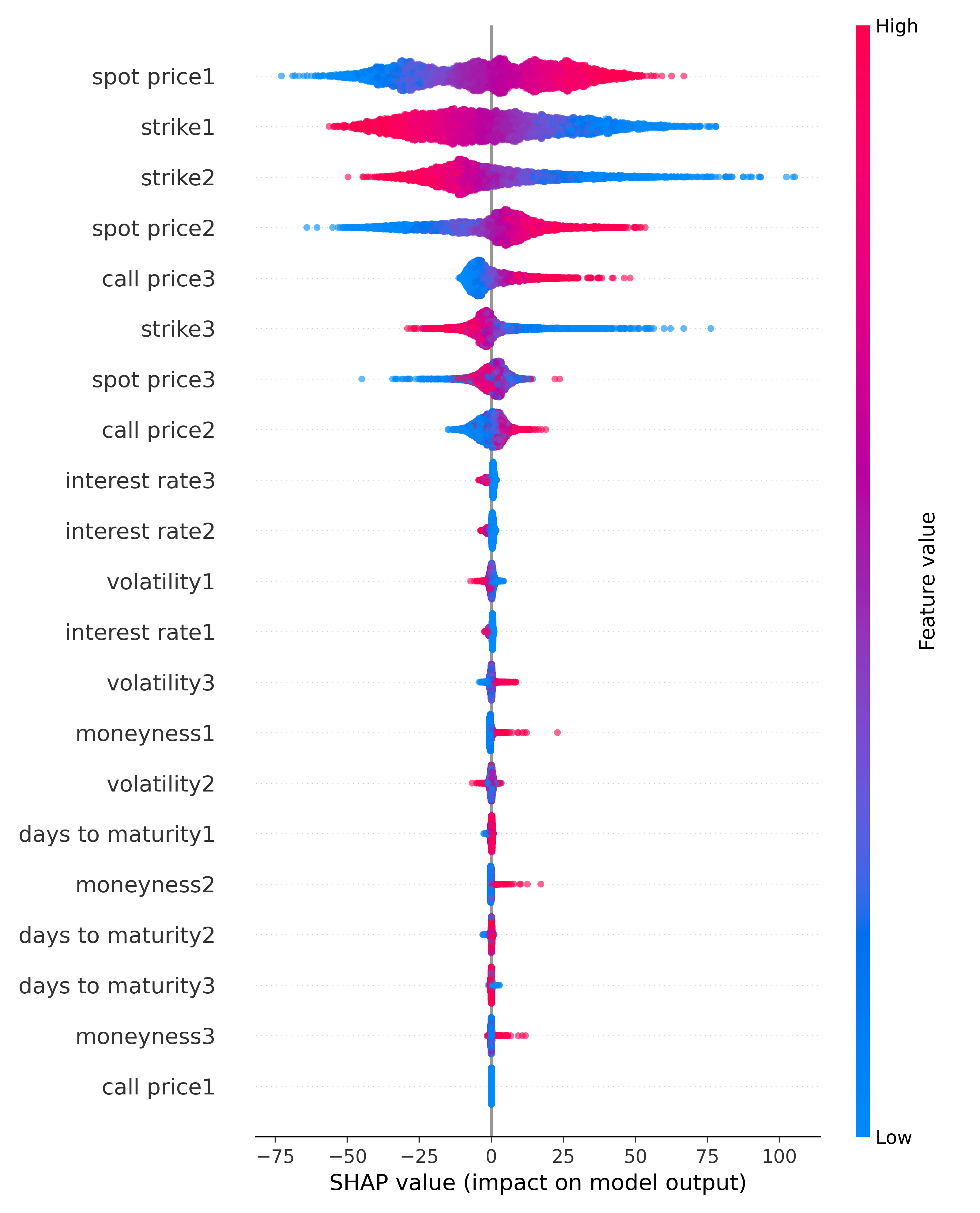}
}
\quad
\subfigure[Moneyness $>$ 1.03 \& 9 days $<$ Maturity $\leq$ 30 days]{
\includegraphics[width=8cm]{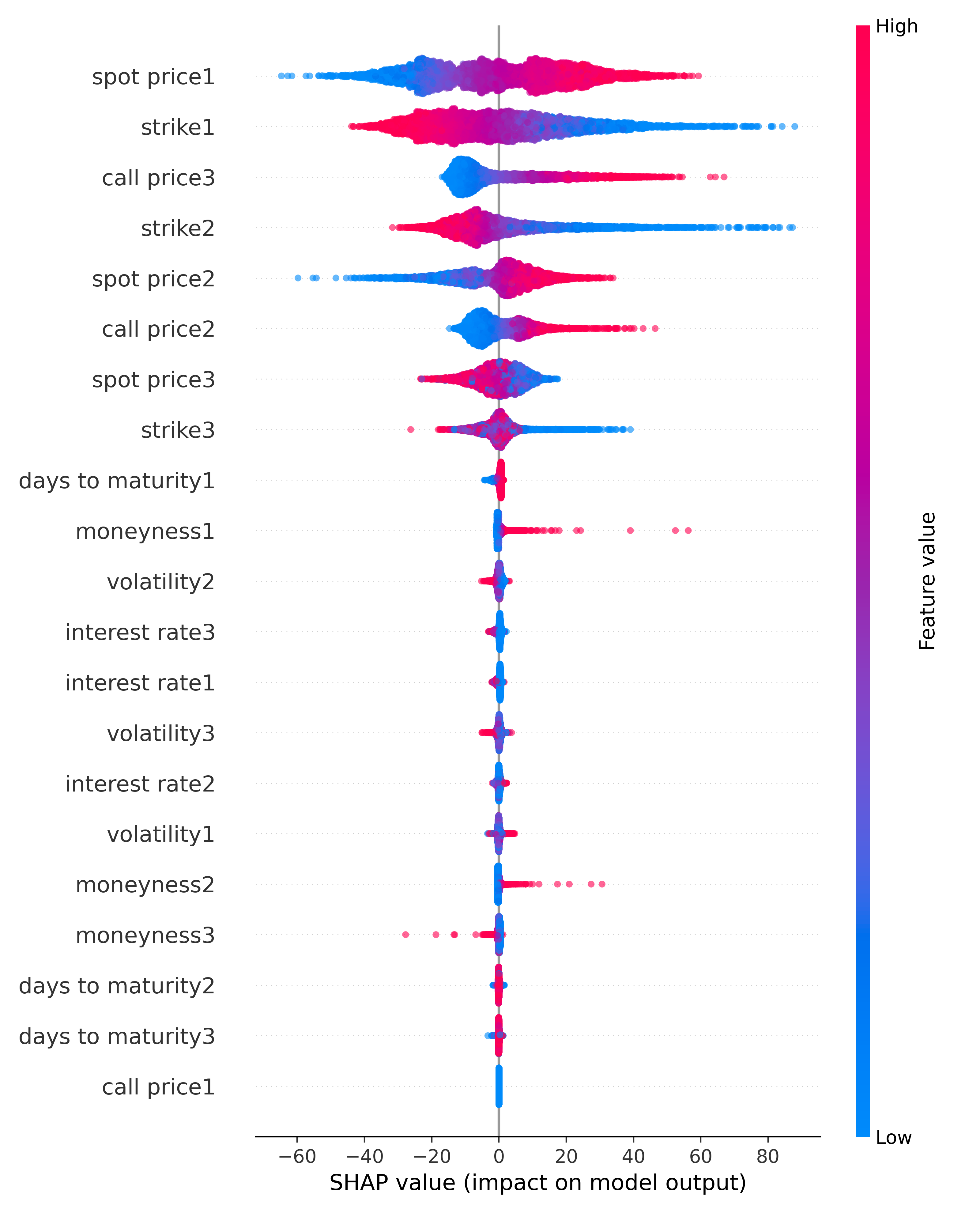}
}
\caption{SHAP plots for options sorted by moneyness and maturity}
\label{SHAP plots for options sorted by moneyness and maturity}
\end{figure*}

\begin{figure*}\ContinuedFloat
\centering
\subfigure[Moneyness $>$ 1.03 \& 30 days $<$ Maturity $\leq$ 90 days]{
\includegraphics[width=8.53cm]{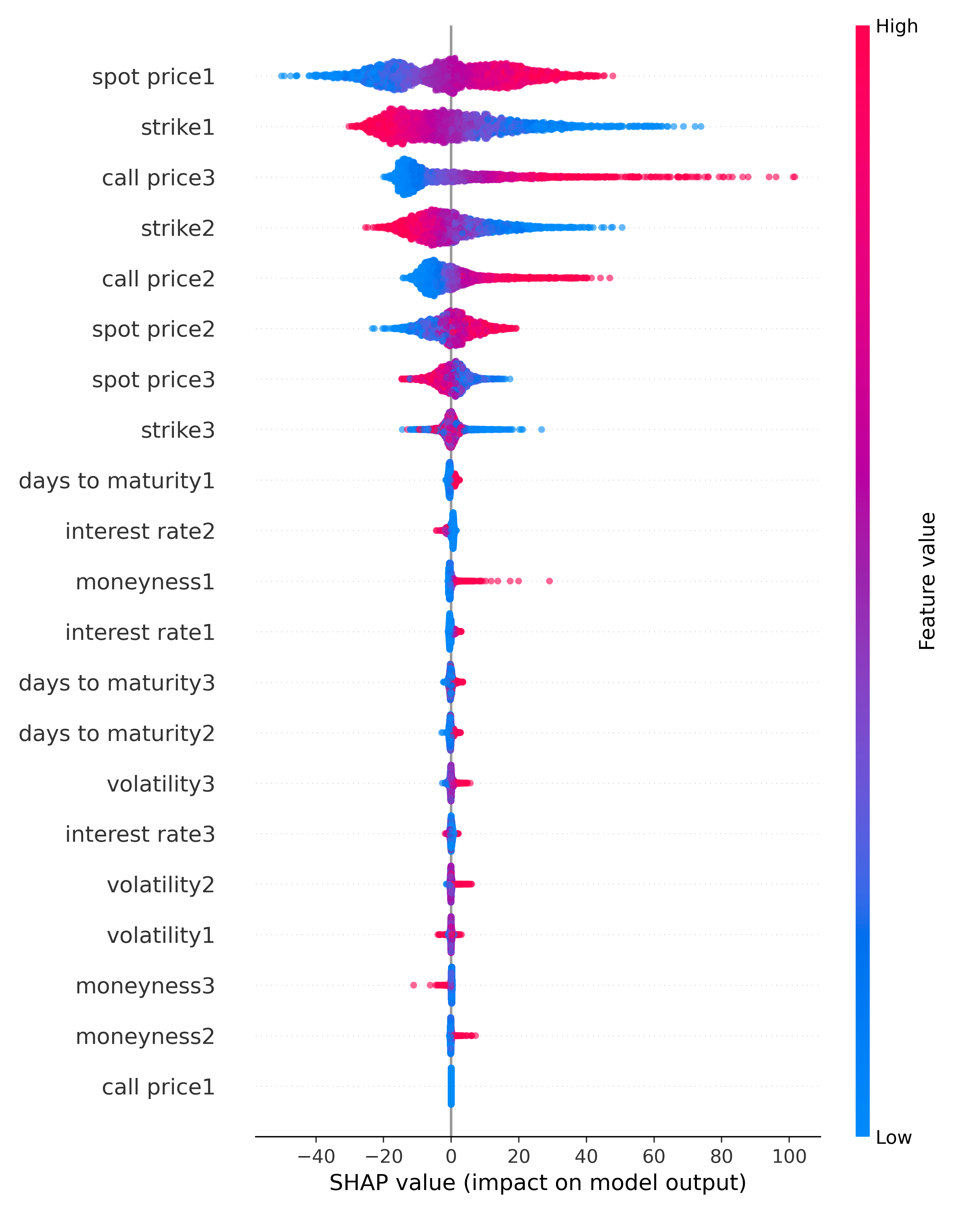}
}
\quad
\subfigure[Moneyness $>$ 1.03 \& 90 days $<$ Maturity $\leq$ 180 days]{
\includegraphics[width=8.53cm]{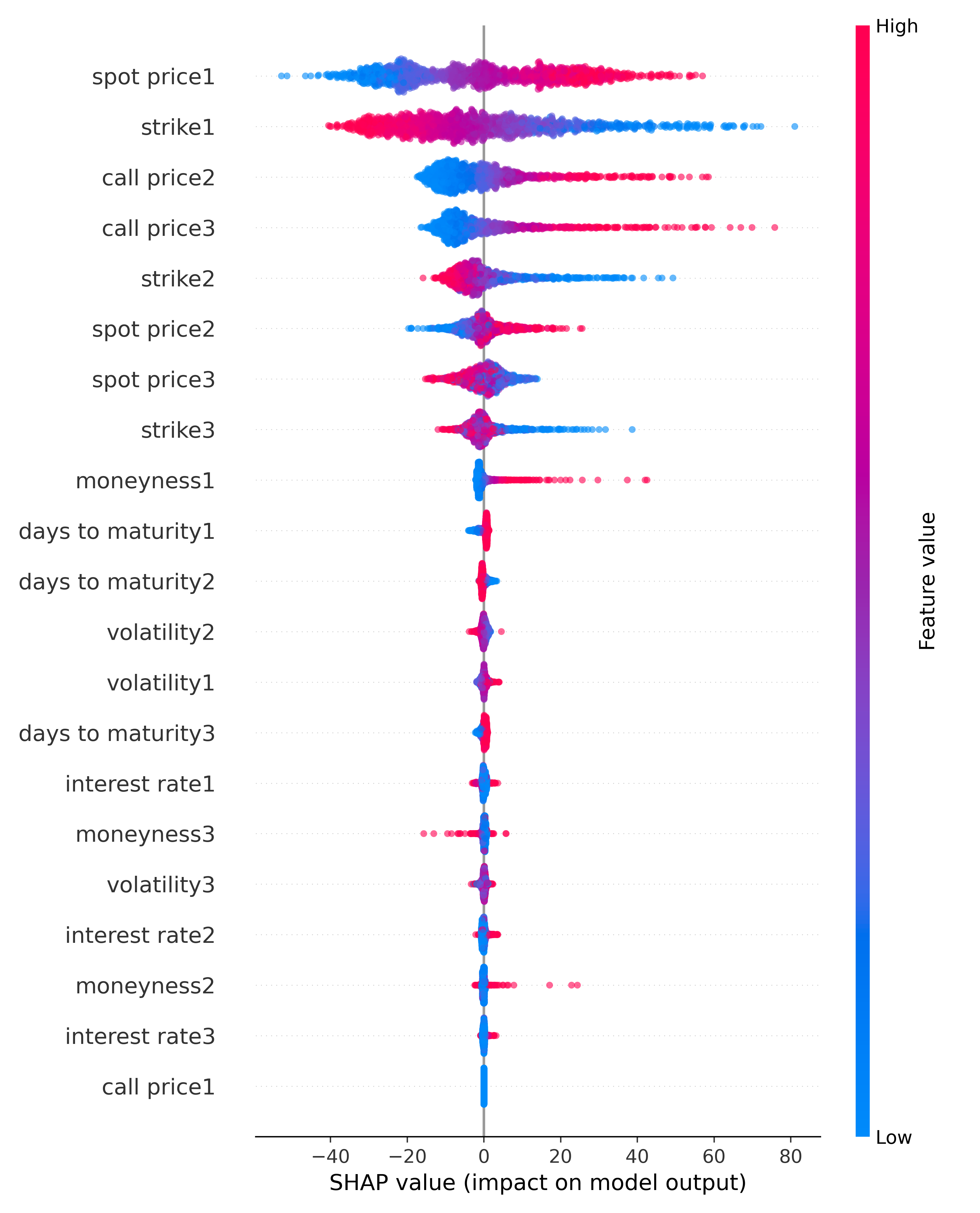}
}
\quad
\subfigure[Moneyness $>$ 1.03 \& Maturity $>$ 180 days]{
\includegraphics[width=8.53cm]{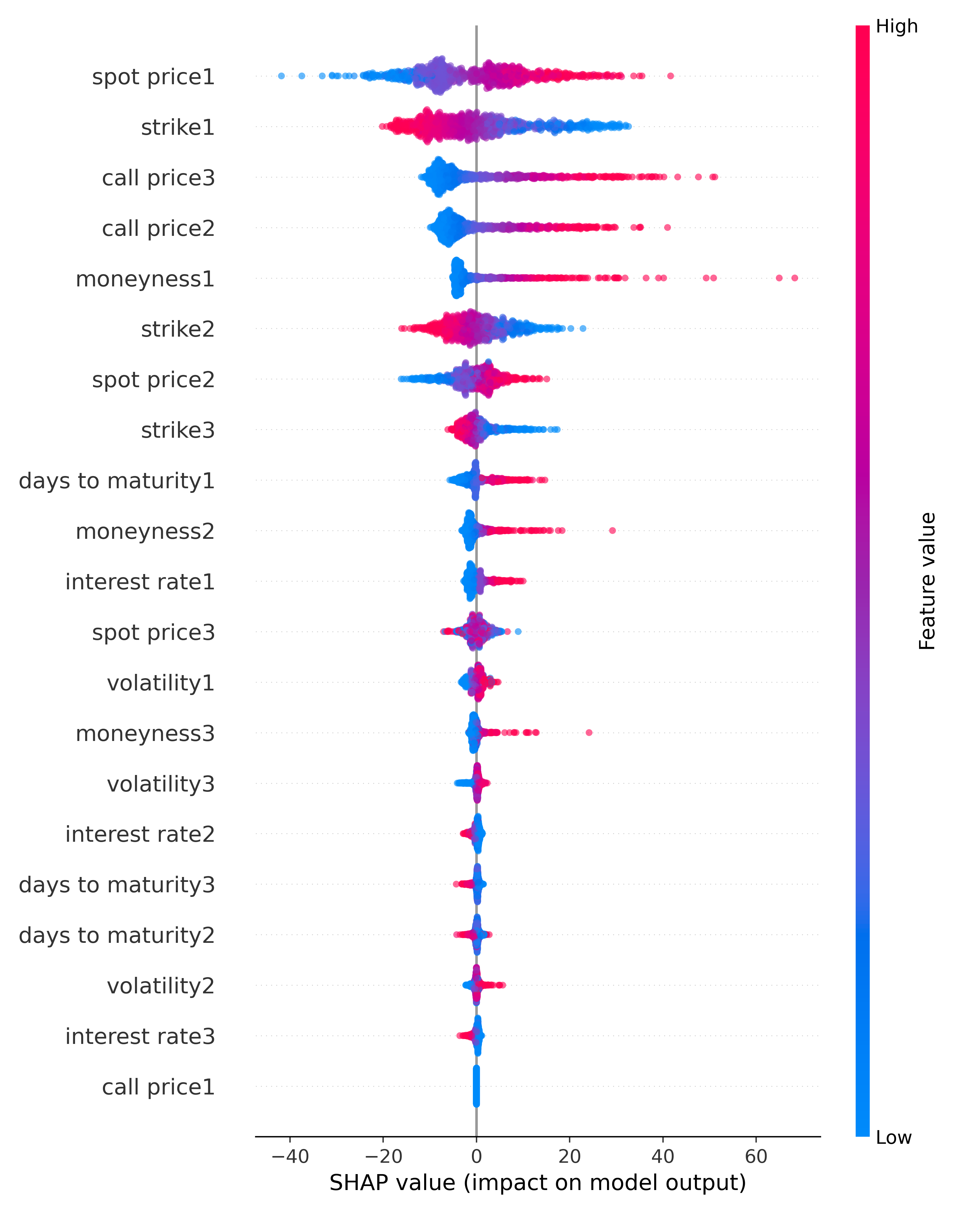}
}
\caption{SHAP plots for options sorted by moneyness and maturity}
\label{SHAP plots for options sorted by moneyness and maturity}
\end{figure*}

\section{Conclusion}
Since the introduction of the renowned Black-Scholes model in 1973, methods for option pricing have undergone extensive research, expansion, and refinement. In contrast to European options, American options, due to their early exercise attribute, are more susceptible to influence from investors, market conditions, and macroeconomic factors. Hence, the study of American option pricing is notably intricate. Nonetheless, the flexibility and variety intrinsic to American options render them favored among investors. As a matter of fact, the vast majority of traded options are of the American type. This inclination underscores the necessity for pricing methodologies that exhibit high precision and computational efficiency, in order to prevent arbitrage opportunities from arising in the financial market.

Starting from the 2010s, the rapid advancement of technologies such as deep learning, big data, and cloud computing has resulted in significant breakthroughs within artificial intelligence. These advances have permeated multiple fields including natural language processing, image recognition, speech recognition, and autonomous driving. The financial sector has also witnessed a surge in research applying artificial intelligence for auxiliary analysis. This paper focuses specifically on the pivotal field of option pricing and amalgamates it with state-of-the-art deep learning techniques. Deep learning models offer distinct advantages in handling non-linear models and capturing temporal relationships. Additionally, they demonstrate considerable improvements in computational efficiency and operational viability compared to traditional models for pricing American options. Furthermore, this study first employs the application of SHapley Additive exPlanations (SHAP), a machine learning interpretability approach, to elucidate the pricing of American options. This addresses a previous limitation of machine learning – its "black-box" nature, which can be challenging to comprehend.

The dataset chosen for this study covers the transaction records of SPY call options from January 2020 to December 2022. A comparative analysis is conducted between the classic binomial tree model and deep learning models such as MLP, LSTM, GRU, and self-attention mechanism. Empirical experiments substantiate that deep learning models present a precision enhancement of over 50 times compared to the traditional model, while requiring less computational time. When classifying options by moneyness, the binomial tree model provides the most accurate predictions for ITM options and the least accurate predictions for OTM options. 
Among the deep learning models, the fundamental MLP model significantly improves the accuracy of OTM option prices.
Subsequent complex deep learning models yield greater accuracy improvements for ITM and ATM options than OTM type.
Ultimately, it turns out that the self-attention GRU model with historical features emerges as the top performer.

Through interpretability analysis, the study reveals that, overall, current spot price, strike, and moneyness have the greatest impact on model predictions. However, the contribution of features for ATM options differs slightly from the other two categories. 
For ATM options, moneyness proves a superior model feature in comparison to isolated spot price and strike. 
Incorporating historical data into option price prediction yields superior results compared to disregarding historical data. Moreover, historical call prices show positive impacts on current call price predictions.

In an innovative way, this study matches options of varying maturity with corresponding implied volatility indices and risk-free interest rates to better uncover the influence of volatility and interest rates on option pricing. The findings indicate that current underlying volatility displays a substantial effect on OTM and ATM options, with a relatively smaller impact on ITM options. However, as maturity increases, the volatility tends to increasingly have a favorable effect on ITM options. The impact of historical volatility is relatively minor and negative overall (with the greatest effect on ATM options). 
The consistent impact of current interest rates is negative for options with shorter days to maturity and positive for options with longer days to maturity, although the impacts are very small in contrast to other inputs.

Employing deep learning techniques to investigate American option pricing infuses this field with fresh vigor and limitless potential. 
Future research directions include the application of deep learning models in other aspects of option pricing, such as forecasting volatility surfaces and Greek letters. 
Given the complexities in gathering historical data for this study, we can try alternative extraction methods to obtain longer time series data in order to optimize the utilization of self-attention in extended time series. 
Furthermore, the employment of SHAP as a machine learning interpretable method holds promise for enhancing the legitimacy and reliability of applying deep learning models in option pricing.

\bibliographystyle{unsrt}
\bibliography{sample-bibliography}

\begin{thebibliography}{10}

\bibitem{black1973pricing}
Fischer Black and Myron Scholes.
\newblock The pricing of options and corporate liabilities.
\newblock {\em Journal of political economy}, 81(3):637--654, 1973.

\bibitem{cox1979option}
John~C Cox, Stephen~A Ross, and Mark Rubinstein.
\newblock Option pricing: A simplified approach.
\newblock {\em Journal of financial Economics}, 7(3):229--263, 1979.

\bibitem{boyle1986option}
Phelim~P Boyle.
\newblock Option valuation using a tree-jump process.
\newblock {\em International options journal}, 3:7--12, 1986.

\bibitem{kim2016multi}
Young~Shin Kim, Stoyan Stoyanov, Stelozar Rachev, and Frank Fabozzi.
\newblock Multi-purpose binomial model: Fitting all moments to the underlying geometric brownian motion.
\newblock {\em Economics Letters}, 145:225--229, 2016.

\bibitem{brennan1977valuation}
Michael~J Brennan and Eduardo~S Schwartz.
\newblock The valuation of american put options.
\newblock {\em The Journal of Finance}, 32(2):449--462, 1977.

\bibitem{ikonen2004operator}
Samuli Ikonen and Jari Toivanen.
\newblock Operator splitting methods for american option pricing.
\newblock {\em Applied mathematics letters}, 17(7):809--814, 2004.

\bibitem{nielsen2002penalty}
Bj{\o}rn~Fredrik Nielsen, Ola Skavhaug, and Aslak Tveito.
\newblock Penalty and front-fixing methods for the numerical solution of american option problems.
\newblock {\em Journal of Computational Finance}, 5(4):69--98, 2002.

\bibitem{zhao2007compact}
Jichao Zhao, Matt Davison, and Robert~M Corless.
\newblock Compact finite difference method for american option pricing.
\newblock {\em Journal of Computational and Applied Mathematics}, 206(1):306--321, 2007.

\bibitem{bossaerts1989simulation}
Peter Bossaerts.
\newblock Simulation estimators of optimal early exercise.
\newblock {\em Graduate School of Industrial Administration}, 1989.

\bibitem{tilley1993valuing}
James~A Tilley.
\newblock Valuing american options in a path simulation model.
\newblock {\em Transactions of the Society of Actuaries}, 45:499--519, 1993.

\bibitem{broadie1997enhanced}
Mark Broadie, Paul Glasserman, and Gautam Jain.
\newblock Enhanced monte carlo estimates for american option prices.
\newblock {\em Journal of Derivatives}, 5:25--44, 1997.

\bibitem{broadie1997pricing}
Mark Broadie and Paul Glasserman.
\newblock Pricing american-style securities using simulation.
\newblock {\em Journal of economic dynamics and control}, 21(8-9):1323--1352, 1997.

\bibitem{longstaff2001valuing}
Francis~A Longstaff and Eduardo~S Schwartz.
\newblock Valuing american options by simulation: a simple least-squares approach.
\newblock {\em The review of financial studies}, 14(1):113--147, 2001.

\bibitem{mcculloch1943logical}
Warren~S McCulloch and Walter Pitts.
\newblock A logical calculus of the ideas immanent in nervous activity.
\newblock {\em The bulletin of mathematical biophysics}, 5:115--133, 1943.

\bibitem{malliaris1993neural}
Mary Malliaris and Linda Salchenberger.
\newblock A neural network model for estimating option prices.
\newblock {\em Applied Intelligence}, 3:193--206, 1993.

\bibitem{hutchinson1994nonparametric}
James~M Hutchinson, Andrew~W Lo, and Tomaso Poggio.
\newblock A nonparametric approach to pricing and hedging derivative securities via learning networks.
\newblock {\em The journal of Finance}, 49(3):851--889, 1994.

\bibitem{lajbcygier1997improved}
Paul~R Lajbcygier and Jerome~T Connor.
\newblock Improved option pricing using artificial neural networks and bootstrap methods.
\newblock {\em International journal of neural systems}, 8(04):457--471, 1997.

\bibitem{anders1998improving}
Ulrich Anders, Olaf Korn, and Christian Schmitt.
\newblock Improving the pricing of options: A neural network approach.
\newblock {\em Journal of forecasting}, 17(5-6):369--388, 1998.

\bibitem{bennell2004black}
Julia Bennell and Charles Sutcliffe.
\newblock Black--scholes versus artificial neural networks in pricing ftse 100 options.
\newblock {\em Intelligent Systems in Accounting, Finance \& Management: International Journal}, 12(4):243--260, 2004.

\bibitem{liang2009improving}
Xun Liang, Haisheng Zhang, Jianguo Xiao, and Ying Chen.
\newblock Improving option price forecasts with neural networks and support vector regressions.
\newblock {\em Neurocomputing}, 72(13-15):3055--3065, 2009.

\bibitem{wang2009nonlinear}
Yi-Hsien Wang.
\newblock Nonlinear neural network forecasting model for stock index option price: Hybrid gjr--garch approach.
\newblock {\em Expert Systems with Applications}, 36(1):564--570, 2009.

\bibitem{garcia2000pricing}
Ren{\'e} Garcia and Ramazan Gen{\c{c}}ay.
\newblock Pricing and hedging derivative securities with neural networks and a homogeneity hint.
\newblock {\em Journal of Econometrics}, 94(1-2):93--115, 2000.

\bibitem{park2014parametric}
Hyejin Park, Namhyoung Kim, and Jaewook Lee.
\newblock Parametric models and non-parametric machine learning models for predicting option prices: Empirical comparison study over kospi 200 index options.
\newblock {\em Expert Systems with Applications}, 41(11):5227--5237, 2014.

\bibitem{kelly1994valuing}
David~L Kelly, Jamsheed Shorish, et~al.
\newblock Valuing and hedging american put options using neural networks.
\newblock {\em Unpublished manuscript, Carnegie Mellon University}, 1994.

\bibitem{keber1999option}
Christian Keber.
\newblock Option valuation with the genetic programming approach.
\newblock {\em Computational finance}, 1999:689--703, 1999.

\bibitem{morelli2004pricing}
Marco~J Morelli, Guido Montagna, Oreste Nicrosini, Michele Treccani, Marco Farina, and Paolo Amato.
\newblock Pricing financial derivatives with neural networks.
\newblock {\em Physica A: Statistical Mechanics and its Applications}, 338(1-2):160--165, 2004.

\bibitem{pires2004american}
Michael~Maio Pires and Tshilidzi Marwala.
\newblock American option pricing using multi-layer perceptron and support vector machine.
\newblock In {\em 2004 IEEE International Conference on Systems, Man and Cybernetics (IEEE Cat. No. 04CH37583)}, volume~2, pages 1279--1285. IEEE, 2004.

\bibitem{jang2019generative}
Huisu Jang and Jaewook Lee.
\newblock Generative bayesian neural network model for risk-neutral pricing of american index options.
\newblock {\em Quantitative Finance}, 19(4):587--603, 2019.

\bibitem{kohler2010pricing}
Michael Kohler, Adam Krzy{\.z}ak, and Nebojsa Todorovic.
\newblock Pricing of high-dimensional american options by neural networks.
\newblock {\em Mathematical Finance: An International Journal of Mathematics, Statistics and Financial Economics}, 20(3):383--410, 2010.

\bibitem{siami2018comparison}
Sima Siami-Namini, Neda Tavakoli, and Akbar~Siami Namin.
\newblock A comparison of arima and lstm in forecasting time series.
\newblock In {\em 2018 17th IEEE international conference on machine learning and applications (ICMLA)}, pages 1394--1401. IEEE, 2018.

\bibitem{hiransha2018nse}
Ma~Hiransha, E~Ab Gopalakrishnan, Vijay~Krishna Menon, and KP~Soman.
\newblock Nse stock market prediction using deep-learning models.
\newblock {\em Procedia computer science}, 132:1351--1362, 2018.

\bibitem{hirsa2019supervised}
Ali Hirsa, Tugce Karatas, and Amir Oskoui.
\newblock Supervised deep neural networks (dnns) for pricing/calibration of vanilla/exotic options under various different processes.
\newblock {\em arXiv preprint arXiv:1902.05810}, 2019.

\bibitem{livieris2020cnn}
Ioannis~E Livieris, Emmanuel Pintelas, and Panagiotis Pintelas.
\newblock A cnn--lstm model for gold price time-series forecasting.
\newblock {\em Neural computing and applications}, 32:17351--17360, 2020.

\bibitem{becker2019deep}
Sebastian Becker, Patrick Cheridito, and Arnulf Jentzen.
\newblock Deep optimal stopping.
\newblock {\em The Journal of Machine Learning Research}, 20(1):2712--2736, 2019.

\bibitem{shapley1953value}
Lloyd~S Shapley et~al.
\newblock A value for n-person games.
\newblock 1953.

\bibitem{lundberg2017unified}
Scott~M Lundberg and Su-In Lee.
\newblock A unified approach to interpreting model predictions.
\newblock {\em Advances in neural information processing systems}, 30, 2017.

\bibitem{merton1973theory}
Robert~C Merton.
\newblock Theory of rational option pricing.
\newblock {\em The Bell Journal of economics and management science}, pages 141--183, 1973.

\bibitem{crank1947practical}
John Crank and Phyllis Nicolson.
\newblock A practical method for numerical evaluation of solutions of partial differential equations of the heat-conduction type.
\newblock In {\em Mathematical proceedings of the Cambridge philosophical society}, volume~43, pages 50--67. Cambridge University Press, 1947.

\bibitem{wanner1996solving}
Gerhard Wanner and Ernst Hairer.
\newblock {\em Solving ordinary differential equations II}, volume 375.
\newblock Springer Berlin Heidelberg New York, 1996.

\bibitem{yousuf2023partial}
Muhammad Yousuf and Abdul~QM Khaliq.
\newblock Partial differential integral equation model for pricing american option under multi state regime switching with jumps.
\newblock {\em Numerical Methods for Partial Differential Equations}, 39(2):890--912, 2023.

\bibitem{broadie1997sotchastic}
Mark Broadie, Paul Glasserman, et~al.
\newblock A sotchastic mesh method for pricing high-dimensional american options.
\newblock Technical report, 1997.

\bibitem{ivașcu2021option}
Codruț-Florin Ivașcu.
\newblock Option pricing using machine learning.
\newblock {\em Expert Systems with Applications}, 163:113799, 2021.

\bibitem{corrado1996skewness}
Charles~J Corrado and Tie Su.
\newblock Skewness and kurtosis in s\&p 500 index returns implied by option prices.
\newblock {\em Journal of Financial research}, 19(2):175--192, 1996.

\bibitem{li2009learning}
Yuxi Li, Csaba Szepesvari, and Dale Schuurmans.
\newblock Learning exercise policies for american options.
\newblock In {\em Artificial intelligence and statistics}, pages 352--359. PMLR, 2009.

\bibitem{lin2021american}
Jingying Lin and Caio Almeida.
\newblock American option pricing with machine learning: An extension of the longstaff-schwartz method.
\newblock {\em Brazilian Review of Finance}, 19(3):85--109, 2021.

\bibitem{gaspar2020neural}
Raquel~M Gaspar, Sara~D Lopes, and Bernardo Sequeira.
\newblock Neural network pricing of american put options.
\newblock {\em Risks}, 8(3):73, 2020.

\bibitem{li2022iteration}
Nan Li.
\newblock An iteration algorithm for american options pricing based on reinforcement learning.
\newblock {\em Symmetry}, 14(7):1324, 2022.

\bibitem{anderson2022accelerated}
David Anderson and Urban Ulrych.
\newblock Accelerated american option pricing with deep neural networks.
\newblock {\em Swiss Finance Institute Research Paper}, (22-03), 2022.

\bibitem{heston1993closed}
Steven~L Heston.
\newblock A closed-form solution for options with stochastic volatility with applications to bond and currency options.
\newblock {\em The review of financial studies}, 6(2):327--343, 1993.

\bibitem{hochreiter1997long}
Sepp Hochreiter and J{\"u}rgen Schmidhuber.
\newblock Long short-term memory.
\newblock {\em Neural computation}, 9(8):1735--1780, 1997.

\bibitem{liang2020forecasting}
Longyue Liang and Xuanye Cai.
\newblock Forecasting peer-to-peer platform default rate with lstm neural network.
\newblock {\em Electronic Commerce Research and Applications}, 43:100997, 2020.

\bibitem{chen2021numerical}
Yinghao Chen, Hanyu Yu, Xiangyu Meng, Xiaoliang Xie, Muzhou Hou, and Julien Chevallier.
\newblock Numerical solving of the generalized black-scholes differential equation using laguerre neural network.
\newblock {\em Digital Signal Processing}, 112:103003, 2021.

\bibitem{atsalakis2009surveying}
George~S Atsalakis and Kimon~P Valavanis.
\newblock Surveying stock market forecasting techniques--part ii: Soft computing methods.
\newblock {\em Expert Systems with applications}, 36(3):5932--5941, 2009.

\bibitem{karagozoglu2022option}
Ahmet~K Karagozoglu.
\newblock Option pricing models: From black-scholes-merton to present.
\newblock {\em The Journal of Derivatives}, 2022.

\bibitem{ruf2019neural}
Johannes Ruf and Weiguan Wang.
\newblock Neural networks for option pricing and hedging: a literature review.
\newblock {\em arXiv preprint arXiv:1911.05620}, 2019.

\bibitem{cavalcante2016computational}
Rodolfo~C Cavalcante, Rodrigo~C Brasileiro, Victor~LF Souza, Jarley~P Nobrega, and Adriano~LI Oliveira.
\newblock Computational intelligence and financial markets: A survey and future directions.
\newblock {\em Expert Systems with Applications}, 55:194--211, 2016.

\bibitem{tkavc2016artificial}
Michal Tk{\'a}{\v{c}} and Robert Verner.
\newblock Artificial neural networks in business: Two decades of research.
\newblock {\em Applied Soft Computing}, 38:788--804, 2016.

\bibitem{riyazahmed2021neural}
Dr~K Riyazahmed.
\newblock Neural networks in finance: A descriptive systematic review.
\newblock {\em Riyazahmed, K (2021). Neural Networks in Finance-A Descriptive Systematic Review. Indian journal of Banking and Finance}, 5(2):1--27, 2021.

\bibitem{zirilli1997financial}
Joseph~S Zirilli.
\newblock {\em Financial prediction using neural networks}, volume 254.
\newblock International Thomson Computer Press UK, 1997.

\bibitem{nagel2021machine}
Stefan Nagel.
\newblock {\em Machine learning in asset pricing}, volume~8.
\newblock Princeton University Press, 2021.

\bibitem{hull2021machine}
John~C Hull.
\newblock Machine learning in business: An introduction to the world of data science.
\newblock {\em (No Title)}, 2021.

\bibitem{pasquale2015black}
Frank Pasquale.
\newblock {\em The black box society: The secret algorithms that control money and information}.
\newblock Harvard University Press, 2015.

\bibitem{doshi2017accountability}
Finale Doshi-Velez, Mason Kortz, Ryan Budish, Chris Bavitz, Sam Gershman, David O'Brien, Kate Scott, Stuart Schieber, James Waldo, David Weinberger, et~al.
\newblock Accountability of ai under the law: The role of explanation.
\newblock {\em arXiv preprint arXiv:1711.01134}, 2017.

\bibitem{bibal2016interpretability}
Adrien Bibal and Beno{\^\i}t Fr{\'e}nay.
\newblock Interpretability of machine learning models and representations: an introduction.
\newblock In {\em 24th european symposium on artificial neural networks, computational intelligence and machine learning}, pages 77--82. CIACO, 2016.

\bibitem{lipton2018mythos}
Zachary~C Lipton.
\newblock The mythos of model interpretability: In machine learning, the concept of interpretability is both important and slippery.
\newblock {\em Queue}, 16(3):31--57, 2018.

\bibitem{friedman2001greedy}
Jerome~H Friedman.
\newblock Greedy function approximation: a gradient boosting machine.
\newblock {\em Annals of statistics}, pages 1189--1232, 2001.

\bibitem{zhu2019forecasting}
You Zhu, Li~Zhou, Chi Xie, Gang-Jin Wang, and Truong~V Nguyen.
\newblock Forecasting smes' credit risk in supply chain finance with an enhanced hybrid ensemble machine learning approach.
\newblock {\em International Journal of Production Economics}, 211:22--33, 2019.

\bibitem{ozgur2021machine}
Onder Ozgur, Erdal~Tanas Karagol, and Fatih~Cemil Ozbugday.
\newblock Machine learning approach to drivers of bank lending: evidence from an emerging economy.
\newblock {\em Financial Innovation}, 7:1--29, 2021.

\bibitem{molnar2020limitations}
Christoph Molnar, S~Gruber, and P~Kopper.
\newblock Limitations of interpretable machine learning methods, 2020.

\bibitem{liang2022time}
Longyue Liang and Xuanye Cai.
\newblock Time-sequencing european options and pricing with deep learning--analyzing based on interpretable ale method.
\newblock {\em Expert Systems with Applications}, 187:115951, 2022.

\bibitem{christensen2021machine}
Kim Christensen, Mathias Siggaard, and Bezirgen Veliyev.
\newblock A machine learning approach to volatility forecasting.
\newblock {\em Available at SSRN}, 2021.

\bibitem{ribeiro2016should}
Marco~Tulio Ribeiro, Sameer Singh, and Carlos Guestrin.
\newblock " why should i trust you?" explaining the predictions of any classifier.
\newblock In {\em Proceedings of the 22nd ACM SIGKDD international conference on knowledge discovery and data mining}, pages 1135--1144, 2016.

\bibitem{park2021explainability}
Min~Sue Park, Hwijae Son, Chongseok Hyun, and Hyung~Ju Hwang.
\newblock Explainability of machine learning models for bankruptcy prediction.
\newblock {\em IEEE Access}, 9:124887--124899, 2021.

\bibitem{carta2021explainable}
Salvatore Carta, Alessandro~Sebastian Podda, Diego Reforgiato~Recupero, and Maria~Madalina Stanciu.
\newblock Explainable ai for financial forecasting.
\newblock In {\em International Conference on Machine Learning, Optimization, and Data Science}, pages 51--69. Springer, 2021.

\bibitem{jaeger2021interpretable}
Markus Jaeger, Stephan Kr{\"u}gel, Dimitri Marinelli, Jochen Papenbrock, and Peter Schwendner.
\newblock Interpretable machine learning for diversified portfolio construction.
\newblock {\em The Journal of Financial Data Science}, 2021.

\bibitem{bussmann2021explainable}
Niklas Bussmann, Paolo Giudici, Dimitri Marinelli, and Jochen Papenbrock.
\newblock Explainable machine learning in credit risk management.
\newblock {\em Computational Economics}, 57:203--216, 2021.

\bibitem{misheva2021explainable}
Branka~Hadji Misheva, Joerg Osterrieder, Ali Hirsa, Onkar Kulkarni, and Stephen~Fung Lin.
\newblock Explainable ai in credit risk management.
\newblock {\em arXiv preprint arXiv:2103.00949}, 2021.

\bibitem{gramegna2021shap}
Alex Gramegna and Paolo Giudici.
\newblock Shap and lime: an evaluation of discriminative power in credit risk.
\newblock {\em Frontiers in Artificial Intelligence}, 4:752558, 2021.

\bibitem{moscato2021benchmark}
Vincenzo Moscato, Antonio Picariello, and Giancarlo Sperl{\'\i}.
\newblock A benchmark of machine learning approaches for credit score prediction.
\newblock {\em Expert Systems with Applications}, 165:113986, 2021.

\bibitem{zhao2018american}
Jinsha Zhao.
\newblock American option valuation methods.
\newblock {\em International Journal of Economics and Finance}, 10(5), 2018.

\bibitem{rosenblatt1957perceptron}
Frank Rosenblatt.
\newblock {\em The perceptron, a perceiving and recognizing automaton Project Para}.
\newblock Cornell Aeronautical Laboratory, 1957.

\bibitem{kingma2014adam}
Diederik~P Kingma and Jimmy Ba.
\newblock Adam: A method for stochastic optimization.
\newblock {\em arXiv preprint arXiv:1412.6980}, 2014.

\bibitem{cho2014learning}
Kyunghyun Cho, Bart Van~Merri{\"e}nboer, Caglar Gulcehre, Dzmitry Bahdanau, Fethi Bougares, Holger Schwenk, and Yoshua Bengio.
\newblock Learning phrase representations using rnn encoder-decoder for statistical machine translation.
\newblock {\em arXiv preprint arXiv:1406.1078}, 2014.

\bibitem{bahdanau2014neural}
Dzmitry Bahdanau, Kyunghyun Cho, and Yoshua Bengio.
\newblock Neural machine translation by jointly learning to align and translate.
\newblock {\em arXiv preprint arXiv:1409.0473}, 2014.

\bibitem{vaswani2017attention}
Ashish Vaswani, Noam Shazeer, Niki Parmar, Jakob Uszkoreit, Llion Jones, Aidan~N Gomez, {\L}ukasz Kaiser, and Illia Polosukhin.
\newblock Attention is all you need.
\newblock {\em Advances in neural information processing systems}, 30, 2017.

\bibitem{meissner2001capturing}
Gunter Meissner and Noriko Kawano.
\newblock Capturing the volatility smile of options on high-tech stocks—a combined garch-neural network approach.
\newblock {\em Journal of economics and finance}, 25(3):276--292, 2001.

\bibitem{pande2006new}
Abhishek Pande and Rajendra Sahu.
\newblock A new approach to volatility estimation and option price prediction for dividend paying stocks.
\newblock In {\em WEHIA 2006--1st International Conference on Economic Sciences with Heterogeneous Interacting Agents; 15--17 June 2006, University of Bologna, Italy}. Citeseer, 2006.

\bibitem{malliaris1996using}
Mary Malliaris and Linda Salchenberger.
\newblock Using neural networks to forecast the s\&p 100 implied volatility.
\newblock {\em Neurocomputing}, 10(2):183--195, 1996.

\bibitem{amornwattana2007hybrid}
Sunisa Amornwattana, David Enke, and Cihan~H Dagli.
\newblock A hybrid option pricing model using a neural network for estimating volatility.
\newblock {\em International Journal of General Systems}, 36(5):558--573, 2007.

\bibitem{yao2000option}
Jingtao Yao, Yili Li, and Chew~Lim Tan.
\newblock Option price forecasting using neural networks.
\newblock {\em Omega}, 28(4):455--466, 2000.

\end{thebibliography}
\end{document}